\documentclass[bibyear]{aa}
\usepackage{amsmath,amstext}
\usepackage{natbib,twoopt}
\usepackage[breaklinks=true,colorlinks=true,linkcolor=blue,citecolor=blue,urlcolor=blue]{hyperref}
\usepackage{wrapfig}
\usepackage{amssymb}
\usepackage{wasysym}
\usepackage{float}
\usepackage{ulem}
\usepackage{breakurl}
\usepackage{color}
\usepackage[flushleft]{threeparttable}

\def\msun{\,{M_\odot}}

\def\spose#1{\hbox to 0pt{#1\hss}}
\def\lta{\mathrel{\spose{\lower 3pt\hbox{$\mathchar"218$}}
     \raise 2.0pt\hbox{$\mathchar"13C$}}}
\def\gta{\mathrel{\spose{\lower 3pt\hbox{$\mathchar"218$}}
     \raise 2.0pt\hbox{$\mathchar"13E$}}}
\def\Mh{{M_{\rm BH}}}

\begin{document}

\newenvironment{tablehere}
  {\def\@@captype{table}}
  {}
\newenvironment{figurehere}
  {\def\@@captype{figure}}
  {}
\makeatother

\titlerunning{Chasing the Light}

\title{Chasing the light: Shadowing, collimation, and the super-Eddington growth of infant black holes in JWST broad-line AGNs}

\author{Piero Madau\inst{1,2}}
\institute{ 
Dipartimento di Fisica ``G. Occhialini,'' Università degli Studi di Milano-Bicocca, Piazza della Scienza 3, I-20126 Milano, Italy \and
Department of Astronomy \& Astrophysics, University of California, 1156 High Street, Santa Cruz, CA 95064, USA}

\abstract{
Observations with the James Webb Space Telescope (JWST) have uncovered a substantial population of high-redshift broad-line active galactic nuclei (BLAGNs) characterized by moderate luminosities, weak X-ray emissions, and faint high-ionization lines. We propose that a subset of these BLAGNs, the so-called ``little blue dots'' (LBDs) are accreting at super-Eddington rates and use geometrically thick, non-advective disk models to investigate photon scattering and shadowing within the polar funnel. Our models predict 
extremely blue optical-UV continuum slopes and highly collimated radiation fields where isotropic-equivalent luminosities exceed the Eddington limit in the polar direction, while shadowing suppresses emission at higher inclinations. This ``searchlight'' configuration naturally generates a stratified ionization structure: coronal and high-excitation narrow lines are produced along the symmetry axis, while the equatorial broad-line region (BLR) remains shielded from the hardest ionizing photons. We show that the anisotropic illumination of the BLR explains the observed faintness of high-ionization lines despite strong Balmer emission. For $\Mh=10^{7.5}$--$10^{8}\,\msun$ black holes accreting at Eddington ratios $\sim 10$, standard BLR conditions predict  He{\sc ii}~$ \lambda 4686$/H$\beta$ in the range of 0.08--0.28. Notably, because inherently blue disk spectra provide a much higher ratio of ionizing to optical photons than standard quasar composites, the observed large Balmer equivalent widths are matched with typical BLR covering factors without invoking enshrouded geometries. Taken together, these findings support the view that super-Eddington accretion flows, shaped by thick disk geometries, may naturally account for the ionizing SED and  emission line diagnostics of high-$z$ LBDs, while offering a plausible pathway to rapid black hole growth at cosmic dawn.
}

\keywords{Accretion (14); Active galactic nuclei (16); James Webb Space Telescope (2291); Supermassive black holes (1663)}

\maketitle
\section{Introduction}

Deep surveys with JWST have revealed an emerging population of compact broad-line active galctic nuclei (BLAGNs) at redshifts of $z>4$. Their abundance lies orders of magnitude above a faint-end extrapolation of the quasar luminosity function \citep{Harikane2023AGN,Juod2026}, and they appear to be powered by early massive black holes (MBHs) with $M_{\rm BH}\sim10^6$–$10^8,\msun$ \citep[e.g.,][]{Kocevski2023,Maiolino2025}. These black-hole masses have been reported to lie above expectations from the local $\Mh$--$M_\star$ relation \citep{Pacucci2023,maiolino2024J,Yue2024}, although this inference is sensitive to selection effects and to systematics in both virial $\Mh$ (e.g., line-profile decomposition, non-virial broad components) and host $M_\star$ estimates \citep[e.g.,][]{Ananna2024,Li2025,Bertemes2025,Lupi2024}. Furthermore, BLAGNs display exceptionally weak X-ray emissions \citep[e.g.,][] {Yue2024,Maiolino2025} and faint high-ionization emission lines \citep{Tang2025,Zucchi2025}. Recent work has introduced a convenient shorthand that classifies these compact JWST BLAGNs by their UV–optical continuum slopes into ``little red dots'' (LRDs) and ``little blue dots'' (LBDs)
\citep{Brazzini2026}. The LRDs are the redder, rest-optical V-shaped minority, and display in some cases 
Balmer line absorption features \citep[e.g.,] {Matthee2024,Kocevski2025,Inayoshi2025,deugenio_restabs_2025}, whereas LBDs are the similarly compact but bluer-continuum majority.

A leading hypothesis suggests that accretion at super-Eddington rates may provide insights into several enigmatic features of the newly discovered sources. Early supercritical accretion flows can drive the rapid growth of light black hole seeds during brief accretion episodes \citep{Madau2014,Volonteri2015, Pezzulli2016,Lupi2016}, potentially reducing the reliance on exotic formation pathways for massive seeds during cosmic dawn. This process may also explain the presence of overmassive, ``dormant'' black holes near the epoch of reionization \citep{Juod2024}. In the super-Eddington thick-disk regime, the observed X-ray faintness of JWST BLAGNs could result from intrinsically weak coronal (hard) X-ray emission within a funnel-like reflection geometry \citep{Madau2024} or obscuration at high inclination angles relative to the rotation axis \citep{Pacucci2024}. Under such conditions, a lower black hole mass is needed to generate the same observed luminosity, leading to reduced black-hole-to-stellar mass ratios. Additionally, the anisotropic emission from super-Eddington disks could bias black hole mass estimates, producing artificially high inferred masses or amplifying uncertainties by up to an order of magnitude \citep{Lupi2024}. Spectral energy distribution (SED) models of advection-dominated super-Eddington ``slim disks'' indicate a significant {lack of rest-frame UV continuum emission} \citep[e.g.,][]{Pognan2020}, an effect that could account for the weak high-ionization emission lines observed in JWST active galctic nuclei (AGNs; \citealt{Lambrides2024}). Alternative studies suggest a contrasting view, whereby super-Eddington accretion produces an excess of UV radiation, leading to much bluer continuum slopes \citep{Castello2016,Tang2019,Netzer2019}.

Super-Eddington, non-advective flows around black holes naturally form geometrically thick, $h/r>1$, accretion disks \citep[for a review, see][]{Abramowicz2013}. The shape and luminosity of low-viscosity, rotating, radiation-pressure supported ``tori'' have been computed by several authors under a number of simplifying assumptions \citep[e.g.,][]{Paczynsky1980,Abramowicz1980,Sikora1981,Wiita1982,Wielgus2016}. Such axisymmetric models differ from the advection-dominated slim disk solution with $h/r\lta 0.3$ \citep[e.g.,][]{Abramowicz1988,Sadowski2009,Lasota2016} in that they are radiatively efficient. Numerical hydrodynamic simulations of supercritical accretion flows have shown that the vertical advection of radiation caused by magnetic buoyancy transports energy faster than radiative diffusion, allowing a significant fraction of the photons to escape from the surface of the disk before being trapped and advected into the hole \citep{Jiang2014, Jiang2019}. 
On the smaller scales of black hole X-ray binaries, recent X-ray polarization observations of the supercritical source Cygnus X-3 by IXPE indeed suggest the presence of a narrow funnel that collimates radiation emitted by the accretion flow and obscures the primary source from view \citep{Veledina2024}.

This work aims to utilize the thick disk, non-advective formalism, rather than the underluminous -- for the same accretion rate -- advective slim disk solution, to further investigate the possibility that JWST LBDs at high redshift may be fueled by super-Eddington accretion flows.  We do not attempt to interpret the origin of the red rest-optical continua of LRDs here; rather, we focus on LBDs, for which the observed rest-UV/optical line diagnostics more directly constrain the intrinsic ionizing SED and its inclination dependence.

\section{An approach to supercritical accretion}
\label{sec:PW}

The accretion flows we are interested in are characterized by high, supercritical rates and form geometrically and optically thick tori dominated by radiation pressure. The large thickness of the disk naturally collimates radiation along the rotation axis, producing a super-Eddington photon flux in the evacuated, centrifugally supported funnel region. {The results presented in this section are based on the work of \citet{Paczynsky1980} to which we refer the interested reader for further details.} We adopted a cylindrical coordinate system ($r, \varphi, z)$ centered on a Schwarzschild ($a=0$)  black hole of mass $\Mh$, and used a pseudo-Newtonian potential to mimic general relativistic effects:
\begin{equation}
\Phi=-{G\Mh\over R-r_S};~~~~~R=(r^2+z^2)^{1/2},~~~r_S={2G\Mh\over c^2}.
\end{equation}
Here, $R$ is the spherical radius and $r_S$ is the gravitational (Schwarzschild) radius. The specific angular momentum. $\ell$, depends only on the position coordinate, $r$, in the case of a polytropic gas, and is assumed a priori to be a linear function of the form 

\begin{equation}
\ell(r)=\ell_K(r_{\rm in})+{\cal C}(r-r_{\rm in}).
\label{eq:angmom}
\end{equation}
The intersections of $\ell(r)$ with the Keplerian angular momentum distribution $\ell_K(r)=(G\Mh r)^{1/2}[r/(r-r_S)]$ provide the inner edge of the torus ($r_{\rm in}$), a pressure maximum ($r_c$) located in the equatorial plane within a few gravitational radii, and a transition radius ($r_{\rm out}$) where the torus solution transitions to a thin accretion disk.  Inside $r_c$ the rotation is super-Keplerian ($\ell>\ell_K$), and the excess centrifugal support is balanced by an inward pressure gradient.
On the disk surface the effective potential (gravitational plus centrifugal) is constant. Its value is fixed by requiring that the flow has zero thickness at the inner Keplerian point. The shape of the torus, with half-thickness $z=h(r)$, is then given by
\begin{equation}
h(r)=\left\{\left[{G\Mh(r_{\rm in}-r_S)\over G\Mh-(r_{\rm in}-r_S) J(r)}+r_S\right]^2-r^2\right\}^{1/2},
\end{equation}
where $J(r)\equiv \int_{r_{\rm in}}^r \ell^2(r')dr'/r'^{3}$. 
Rather than assuming a value for the exponent $q$ in $\ell(r)$, one can equivalently specify $r_{\rm out}$. Radiation is emitted from the photosphere at the local Eddington rate,
\begin{equation}
{\vec F}_{\rm rad}=-{c\over \kappa_{\rm es}}{\vec g}_{\rm eff}=
-{c\over \kappa_{\rm es}}\left(-{\vec \nabla}\Phi+{\ell^2\over r^3} {\hat e_r}\right), 
\label{eq:Frad}
\end{equation}
where ${\vec g}_{\rm eff}$ is the effective gravity vector  perpendicular to the surface of the disk and $\kappa_{\rm es}$ is the electron  scattering opacity. The effective temperature on the surface of the disk can be written explicitly as $\sigma T_{\rm eff}^4(r)=F_{\rm rad}=(c/\kappa_{\rm es})\sqrt{Q(r)}$,
with
\begin{equation}
Q(r)\equiv 
{G^2M^2_{\rm BH}\over (R-r_S)^4}+{\ell^4\over r^6}-{2G\Mh\ell^2\over R(R-r_S)^2r^2},
\label{eq:Teff}
\end{equation}
where $R(r)=[r^2 + h(r)^2]^{1/2}$ is the  distance between a point on the disk’s surface and the black hole. %
The total luminosity radiated by the torus, $L_{\rm rad}$, is calculated by integrating the emitted flux over the two disk surfaces:
\begin{equation}
\begin{aligned}
L_{\rm rad} & = 2\int_{0}^{2\pi}\int_{r_{\rm in}}^{r_{\rm out}}F_{\rm rad}\,d\Sigma\\
& = L_{\rm Edd}\int_{r_{\rm in}}^{r_{\rm out}} {rR(R-r_S)^2\over G^2M^2_{\rm BH}h(r)}\,Q(r)dr,
\label{eq:Lrad}
\end{aligned}
\end{equation}
where $d\Sigma=[1+(dh/dr)^2]^{1/2}\,rdrd\varphi$ is the area element. The total luminosity generated by viscosity in the thick portion of the disk can be written as
\begin{equation}
L_{\rm gen}=\dot M\,[(e_{\rm in}-e_{\rm out})
-\Omega_{\rm out}(\ell_{\rm in}-\ell_{\rm out})]
\equiv \varepsilon \dot M c^2,
\label{eq:Lgen}
\end{equation}
where $e=|\Phi|-\ell^2/(2r^2)$ is the specific binding energy, $\Omega=\ell/r^2$ is the angular velocity, and $\dot M$ is the accretion rate. 
Neglecting radial advection, global energy conservation requires that the total energy gain be compensated by radiative losses, $L_{\rm gen}=L_{\rm rad}$. By equating Equations (\ref{eq:Lrad}) and (\ref{eq:Lgen}) one can then calculate the accretion rate. Note that, rather than assuming a value for the constant ${\cal C}$ in the specific angular momentum distribution, in these models one can equivalently specify $\dot M$. Whether the surface conditions described above correspond to a fully realizable 3D equilibrium remains an open question, which is only partially addressed in global magneto-hydrodynamic simulations \citep[e.g.,][]{Jiang2014,Jiang2019,Pacucci2024,Zhang2025}.

\begin{figure}[!htb]
\centering
\includegraphics[width=0.95\hsize]{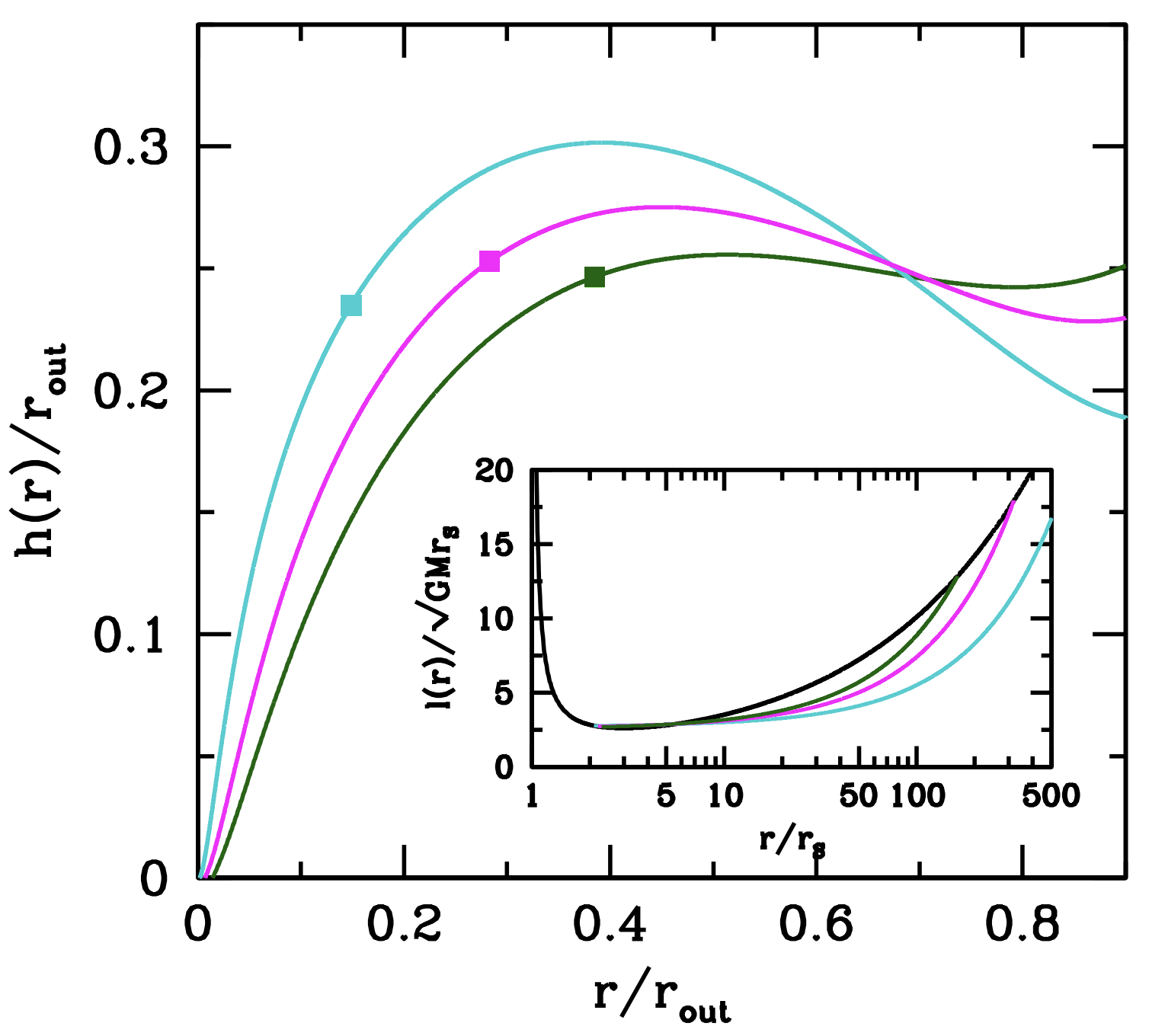}
\caption{Meridional cross sections (over one quadrant) for the supercritical thick disks described by Models A (cyan curve), B (magenta curve), and C (green curve). As $r_{\rm in}$ decreases, $r_{\rm out}$ increases and so  does the ratio $L_{\rm rad}/L_{\rm Edd}$, while  the efficiency of mass to energy conversion becomes progressively smaller. Smaller values of $r_{\rm in}$ also imply steeper and deeper funnels, where pressure gradients are balanced by centrifugal forces rather than by gravity and luminosities can exceed the Eddington limit. The square symbols mark the location on the surface inside which 90\% of the disk luminosity $L_{\rm rad}$ is actually emitted. 
The inset shows the Keplerian specific angular momentum distribution for the adopted pseudo-Newtonian potential (black line) and the three angular momentum distributions corresponding to our supercritical disks (Models A, B, and C).
}
\label{fig:shape}
\end{figure}

\subsection{Constructing  models}

The inset in Figure \ref{fig:shape} shows three representative angular momentum distributions characterizing  moderately supercritical flows around black holes. We considered three values of the inner edge, $r_{\rm in}/r_S=2.1, 2.2,$ and 2.3 (hereafter Models A, B, and C, corresponding to efficiencies of $\varepsilon=0.021,$ 0.035, and 0.044, respectively. These are comparable to the efficiencies observed in  radiation magneto-hydrodynamic simulations of moderately supercritical  flows \citep{Jiang2014,McKinney2014,Sadowski2015,Jiang2019,Huang2023}, and exceeds the efficiencies predicted by slim disk models for these accretion rates \citep[e.g.,][]{Sadowski2009}. 

\begin{figure}[!htb]
\centering
\includegraphics[width=\hsize]{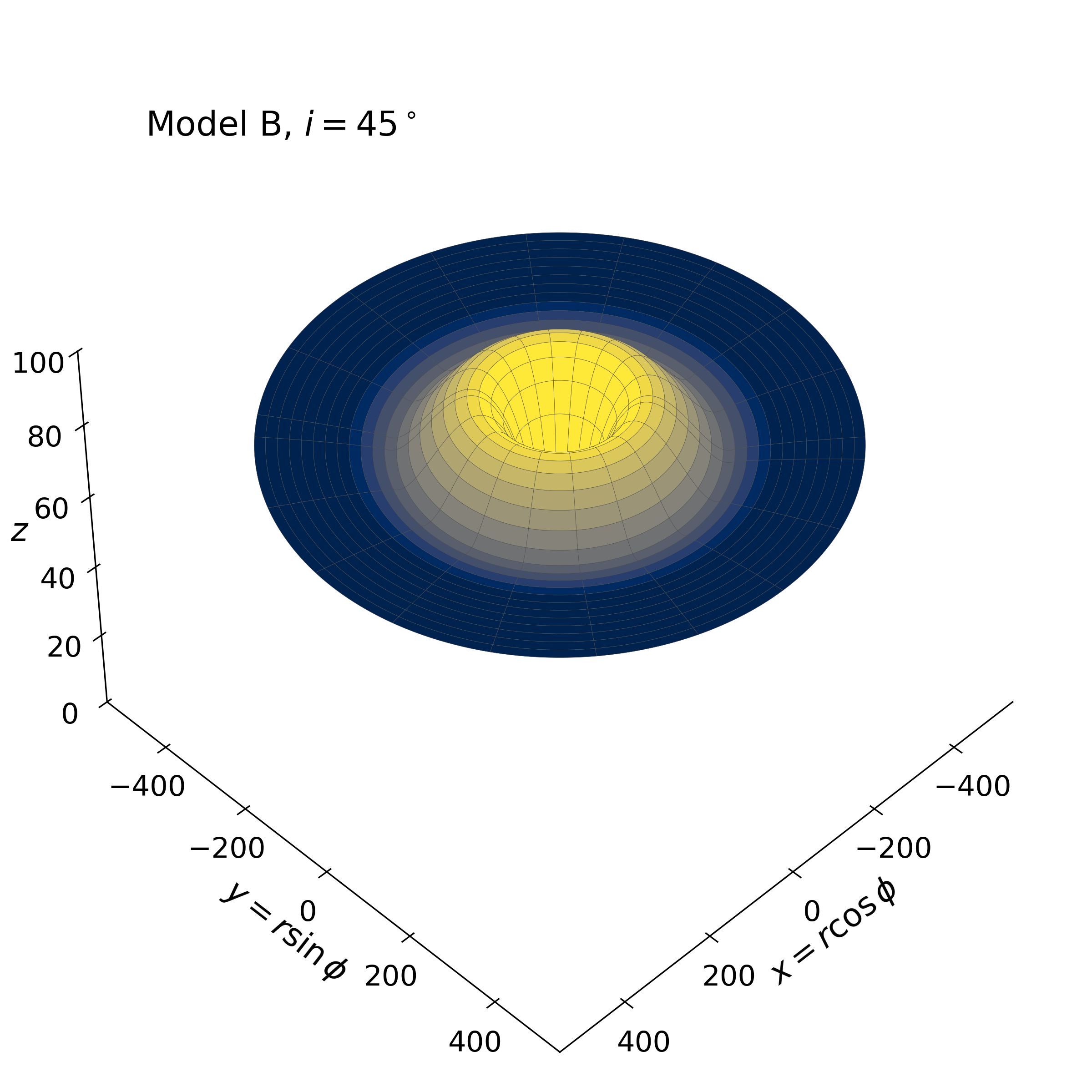}
\includegraphics[width=\hsize]{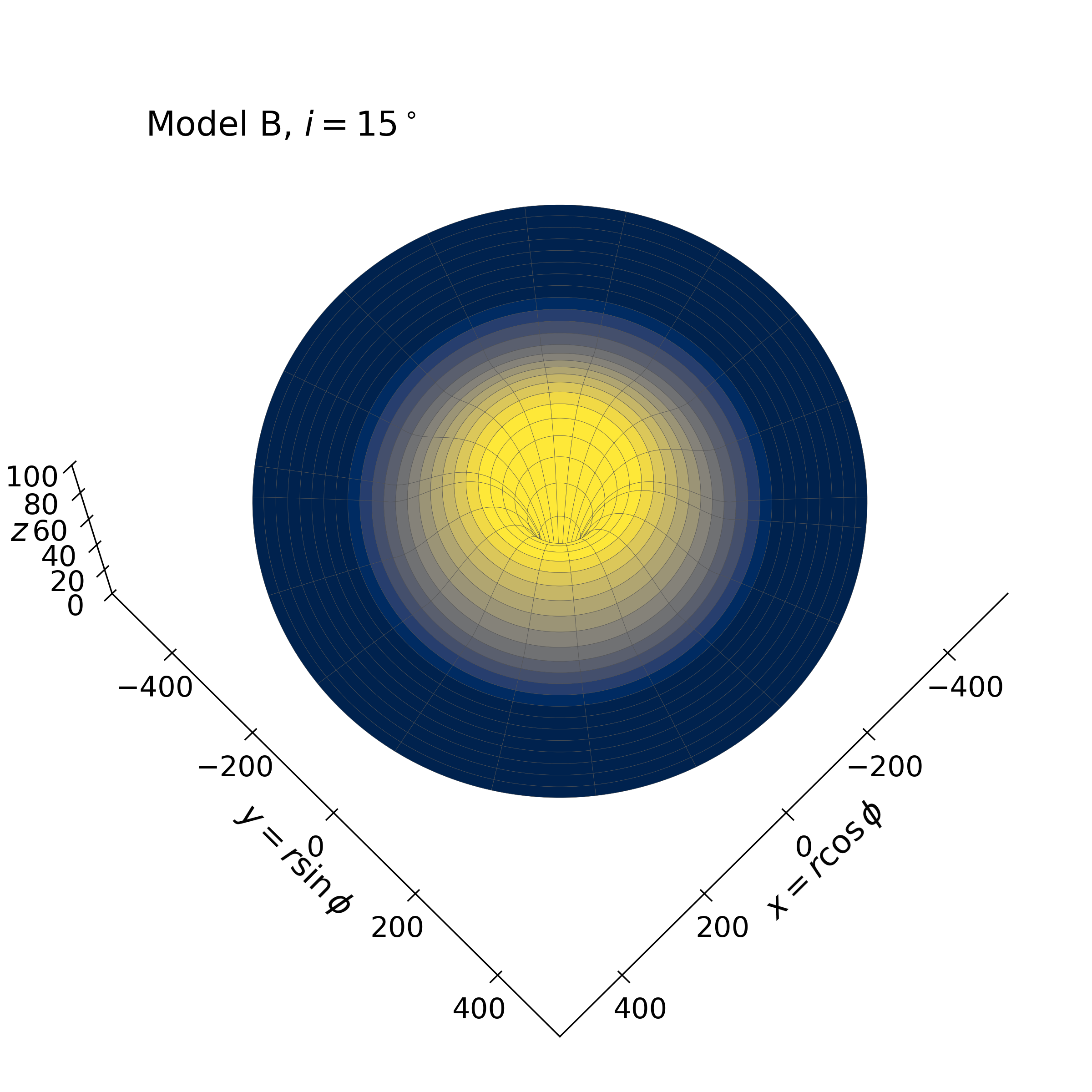}
\caption{
3D toroidal geometry (including shadowing) of a supercritical accretion disk (Model B). The structure is shown at inclination angles $i=45^\circ$ (top) and $i=15^\circ$ (bottom), with $r$ and $z$ expressed in units of $r_S$. The shadow boundary depends on both radius and azimuth (see Appendix~A), and the resulting geometry produces pronounced anisotropy that exceeds the standard inclination–induced anisotropy of a thin accretion disk. The thick funnel extends out to $\sim 300\,r_S$ where it is  matched to an outer geometrically thin disk shown extending to $r=500\,r_S$ to illustrate the geometric transition. The color coding represents the bolometric surface brightness of the accretion disk, where brighter (yellow) colors indicate regions of higher intensity and darker (blue) colors indicate lower intensity.
}
\label{fig:tori}
\end{figure}

The isotropic luminosities radiated by the thick portion of the disk ($r<r_{\rm out}$) in Models A, B, and C are $L_{\rm rad}/L_{\rm Edd}=6.6, 4.3,$ and 3.1, corresponding to $\dot m\equiv \dot M/\dot M_{\rm Edd}=31.7, 12.5,$ and 7.1. Here we have defined the critical accretion rate $\dot M_{\rm Edd} \equiv 10\,L_{\rm Edd}/c^2$ for powering the Eddington luminosity, assuming a 10\% radiative efficiency.  In these models the total radiated  luminosity is only logarithmically dependent on the accretion rate while being linearly proportional to the black hole mass, 
\begin{equation}
L_{\rm rad}/L_{\rm Edd}\simeq 2.26\ln\,\dot m-1.13.
\end{equation}
The shapes of the thick portion of such disks (one quadrant only) are also outlined in Figure \ref{fig:shape}. The narrow, low-density funnels that develop in the innermost regions ($dh/dr>0$, $h/r>1$) can be specified by the opening half-angle, $\Theta$: 
\begin{equation}
\Theta={\rm arccot}\,(h/r)_{\rm max}.
\end{equation}
The three configurations (Models A, B, and C)  have funnel opening half-angles of $\Theta=22^\circ,$ $35^\circ$, and $44^\circ$, respectively. The outer disk is characterized by a flattened region with $dh/dr<0$ and $h/r\lta 1$ that eventually matches a Shakura-Sunyaev thin disk at $r_{\rm out}$. As $r_{\rm in}$ decreases and the accretion flow becomes more super-Eddington, the disk grows larger and fatter, and more of the inner, hot and luminous, funnel regions remain hidden from view, visible only at small viewing angles from the rotation axis of the system. The 3D geometry (including shadowing) of the thick portion of a radiation-supported torus (our Model B) is shown in Figure \ref{fig:tori} for two inclination angles.

We have added a few extra models to our main set to generate  a simple best-fit  ``collimation angle-accretion-rate'' relation for supercritical flows of the form
\begin{equation}
\Theta = 98.5^\circ\,\left({\dot M\over \dot M_{\rm Edd}}\right)^{-0.43}.
\end{equation}
An observer looking down the rotation axis will measure a  highly collimated equivalent isotropic luminosity, which we compute numerically in the next section. The main properties of these super-Eddington disk models are summarized in Table \ref{table:thickdisks}.

It is worth noting that the disk vertical thickness $h(r)$, and hence the funnel opening angle $\Theta$, is set by the assumed specific angular-momentum distribution $\ell(r)$. \citet{Wielgus2016} explored an extremal family of non-advective accretion tori with constant angular momentum ($\ell={\rm const}$), showing that while such models maximize the vertical thickness, the main effect that can reduce collimation in super-Eddington flows is radial advection rather than the detailed form of $\ell(r)$. Nevertheless, global radiation-MHD simulations that include heat advection still produce geometrically thick, $h/r>1$ accretion tori with narrow, low-density funnels along the rotation axis \citep[e.g.,][]{Jiang2019,Pacucci2024,Zhang2025}. 
This phenomenology motivates our use of non-advective analytic torus solutions as an effective description of the photospheric geometry in super-Eddington flows, where narrow polar funnels -- and the associated ``searchlight'' collimation -- are found in simulations that include advective heat transport.

To extend our SED model beyond the outer edge of the super-critical funnel, we matched our thick-disk solution to a standard radiation-pressure-dominated thin disk. We set the transition radius to $r_t=0.9\,r_{\rm out}$, a choice motivated by the need to smoothly connect the geometrical profile of the inflated torus to the outer disk while avoiding boundary effects at the very edge of the numerical grid. In this outer region ($r>r_t$), we calculated the disk vertical profile $h(r)$ assuming hydrostatic equilibrium in the Paczynsky-Wiita potential. Asymptotically, this solution recovers the classical Shakura-Sunyaev regime for a radiation-pressure-dominated disk, characterized by a constant scale height, $h$. For a standard accretion disk where opacity is dominated by electron scattering, this height depends only on the accretion rate,
$h\propto \dot{m}\,r_{\rm S}$ \citep[e.g.,][]{Shakura1973,Paczynsky1980}. We maintained continuity across the transition by matching the effective temperature and radiative flux at $r_t$. While the inner thick disk exhibits a flatter radial temperature profile due to reprocessing and non-Keplerian dynamics, the radiative flux in the thin-disk portion follows the standard Keplerian decline, $F(r)\propto r^{-3}$. As the outer, cooler  thin disk is intrinsically energetically subdominant -- with the integrated luminosity per radial bin falling as $1/r$, it contributes modestly to the UV-optical continuum. This is further reinforced for distant, high-inclination observers by the $\cos i$ projection factor, which suppresses the relative contribution of the planar outer disk compared to the vertically extended inner funnel. %
Finally, we truncated the thin disk at an outer radius of $1500\,r_{\rm S}$, comparable to the self-gravity radius beyond which radiation-pressure-dominated disks are expected to become gravitationally unstable and prone to fragmentation. A representative estimate for this instability radius is \citep[e.g.,][]{Kawaguchi2004}
\begin{equation}
r_{\rm sg}\simeq 10^3\,
\left(\frac{0.1}{\alpha}\right)^{2/9}
\left(\frac{10^{8}\,M_\odot}{M_{\rm BH}}\right)^{-2/9}
\dot{m}^{4/9}\,r_{\rm S},
\end{equation}
ensuring that our model remains within the physically stable regime of the accretion flow.
\begin{table}
  \begin{threeparttable}
    \caption{Properties of thick disks.
\label{table:thickdisks}}
     \begin{tabular}{cc}
        Parameter & Values for Models A, B, and C\\ [2pt]
        \hline
        $r_{\rm in}/r_S$  & 2.1, 2.2, 2.3 \\
        $r_{\rm out}/r_S$ &  1078, 320, 163\\
        $\epsilon$ &  0.021, 0.035, 0.044\\
        $L_{\rm rad}/L_{\rm Edd}$ &  6.6, 4.3, 3.1\\
      
  $\dot m$ &  31.7, 12.4, 7.1\\
        $\Theta$  & 22$^\circ$, 35$^\circ$, 44$^\circ$\\  
        \hline
     \end{tabular}
    \begin{tablenotes}
     \small
      \item {Values are derived from the analytic solutions of \citet{Paczynsky1980}, with notation adapted as described in the text.}
    \end{tablenotes}
  \end{threeparttable}
\end{table}

\subsection{Anisotropic radiation field}

Geometrical thickness causes occultation of the innermost (and hottest) disk regions for high-inclination systems, a shadowing effect that must be taken into account when discussing the observability of super-Eddington flows \citep[e.g.,][]{Sikora1981,Madau1988,Wang2014,Ogawa2017}. An observer at distance $D$ from the disk will observe a bolometric flux,
\begin{equation}
F^i={1\over D^2}\int_{\Sigma_{\rm vis}} I({\hat n})
({\hat  n\cdot \hat N})d\Sigma,
\end{equation}
where ${\hat N}$ is the outward normal to the surface $z=h(r)$, ${\hat n}$ is the direction of the line of sight corresponding to an inclination angle, $i$, with respect to the rotation axis ($\hat n\cdot \hat N>0$ for an area element to be visible to the observer), 
and $\Sigma_{\rm vis}$ is the  emitting surface not occulted by the disk. We assume that light travels in straight lines with no Doppler and gravitational frequency shifts, and that the radiant intensity $I(\hat n)$ is isotropic, $I(\hat n)=F_{\rm rad}/\pi$ for $\hat n\cdot \hat N>0$. And while angle-dependent relativistic effects become more pronounced the closer the emission region is to the hole, their impact on the emergent disk spectrum has been found to be typically negligible for nonspinning black holes \citep[e.g.,][]{Zeltyn2022}. 

\begin{figure}[!htb]
\centering
\includegraphics[width=0.95\hsize]{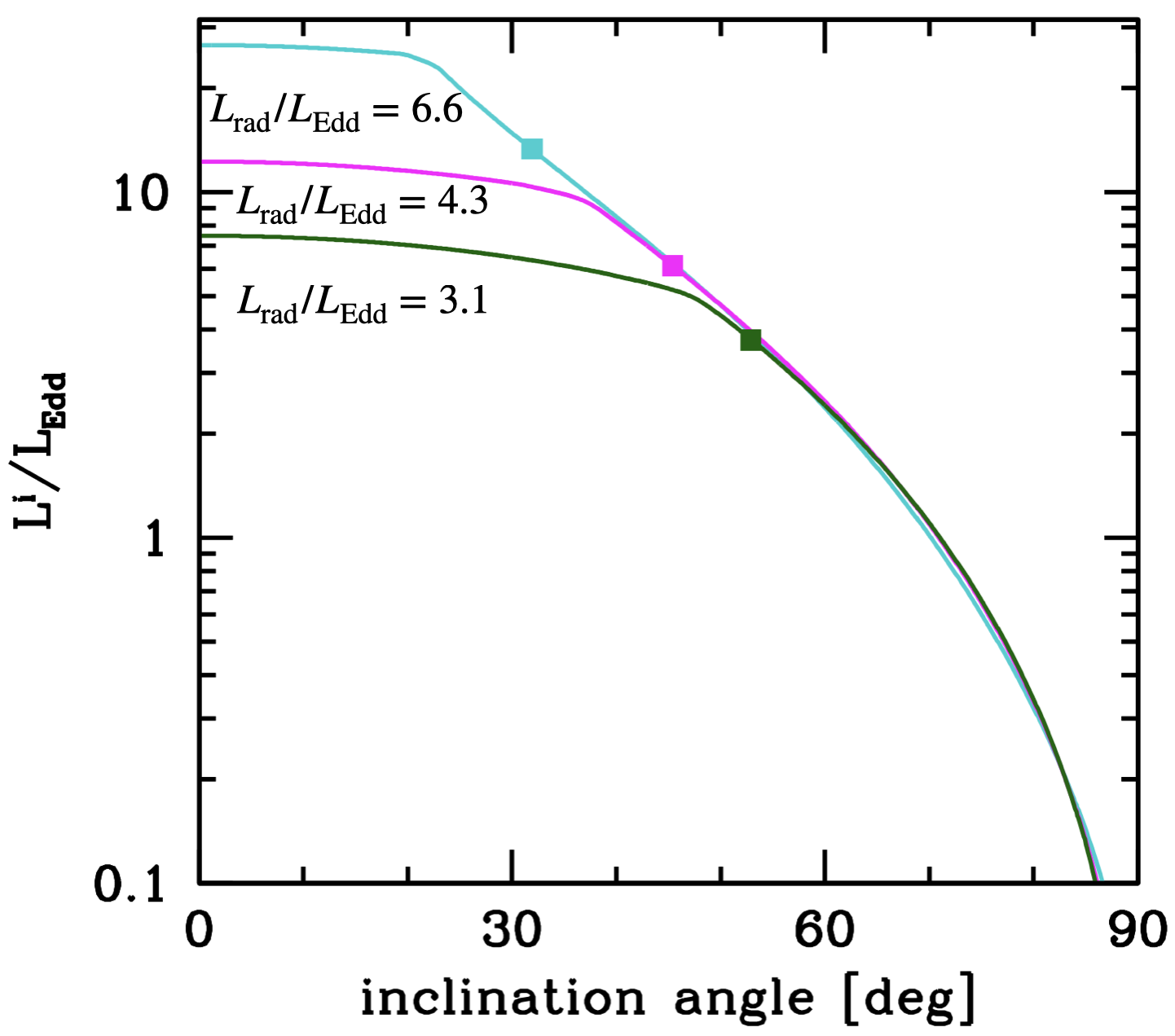}
\caption{Equivalent isotropic luminosity, $L^i$
(in units of $L_{\rm Edd}$), for supercritical disk Models A (turquoise line),  B (magenta line), and C (green line) as measured by an observer at different inclination angles $i$. The labels along the curves mark the true isotropic luminosities (in units of Eddington) radiated by the thick disks. The square symbols mark the half-power beam width --  the inclination angle $i_{1/2}$ at which the observed luminosity falls to 50\% of the peak value. In the wider opening angle case, there is little beam-broadening by scattered radiation. 
}
\label{fig:shadowing}
\end{figure}

In order to get the correct angle-averaged radiation luminosity, one must also account for the fact that photons emitted within the funnel will scatter off the funnel walls many times before escaping either to infinity or falling down into the black hole. In our reflection treatment, photons incident on the funnel wall are re-emitted into a random outward direction while preserving their frequency in the zero-angular-momentum frame. Each area element within the funnel will then see, in addition to the primary locally generated photon flux diffusing from below, incident radiation from neighboring walls that must be included in the balance of forces. Details on how to calculate viewing-angle dependent SEDs including the self-illumination of the inner funnel are given in Appendices A and B. 

Figure \ref{fig:shadowing} shows the resulting ``equivalent isotropic luminosity,'' $L^i \equiv 4\pi D^2\,F^i$, for different inclination angles. Thick disks seen more edge-on, $\Theta> 60^\circ$, appear cooler and underluminous, never exceeding  a few times $L_{\rm Edd}$, regardless of ($\dot M/\dot M_{\rm Edd}$). At smaller inclination angles, the bottom funnel ``hot spot'' becomes more and more visible, and the apparent luminosity increases up to the point where the inclination angle $i$ is approximately equal to the half-opening angle $\Theta$, then flattens for close to pole-on views with
\begin{equation}
{L^{0}\over L_{\rm Edd}}\simeq \left({\dot M\over \dot M_{\rm Edd}}\right).
\end{equation}
An observer looking along the funnel will therefore detect strongly amplified emission. Note how, even in the case of a narrow funnel, the half-power beam width of the thick disk -- the viewing angle, $i_{1/2}$, at which the observed luminosity falls to 50\% of the peak value,  $L^{i_{1/2}}=0.5\,L^{0}$ -- is broader than the half-opening angle, $\Theta$,  because radiation emitted at the bottom of the funnel is scattered sideways  from the upper funnel walls into larger observer angles. Only the outer portions of the disk are observable when viewed edge-on, and these regions are cooler in the case of more supercritical, more extended tori.

\section{Observational implications}

In this section, we explore some observational implications of our findings on super-Eddington accretion flows for JWST-discovered LBDs. 

\subsection{Weak coronal X-ray emission}

Hard X-ray emission from AGNs is believed to result from the thermal Comptonization of optical-UV disk photons by hot electrons in an inner, tenuous corona \citep[e.g.,][]{Haardt1991}. In \citet[][]{Madau2024}, we showed that, in the funnel-like reflection geometry that characterizes supercritical accretion, the nearly isotropic UV photon field will Compton cool coronal plasma to much lower temperatures than in the standard open geometry of a thin accretion disk, and the emerging X-ray continuum is expected to be extremely soft and X-ray weak, with $\alpha_{\rm ox}\equiv 0.384\,\log_{10}\!\left[L_\nu(2\,{\rm keV})/L_\nu(2500\,\text{\AA})\right]\lesssim -1.8$.
Numerical simulations indicate that, while a thin coronal  plasma with gas temperatures $\gta 10^8\,$K is generated in the inner regions of sub-Eddington accretion disks, the fraction of dissipation in this hot component  decreases as the mass accretion rate increases \citep{Jiang2019sE}. 

\subsection{Orientation-dependent quasi-thermal spectra}

At the temperatures and densities of radiation-supported tori, electron scattering dominates absorption as a source of opacity, and the emitted spectrum cannot be approximated as a sum of blackbodies. For the form of the locally produced radiated specific intensity, $I^{\rm local}_\nu(r)$, we adopt here the modified blackbody spectrum appropriate for isotropic scattering in a constant density atmosphere \citep{Rybicki1986,Czerny1987}. Our surface–reflection treatment intentionally adopts the idealized limit in which incident photons are re-emitted into a random outward direction while preserving their frequency in the zero–angular-momentum frame. Details of the calculations are given in Appendix C.

\begin{figure}[!htb]
\centering
\includegraphics[width=\hsize]{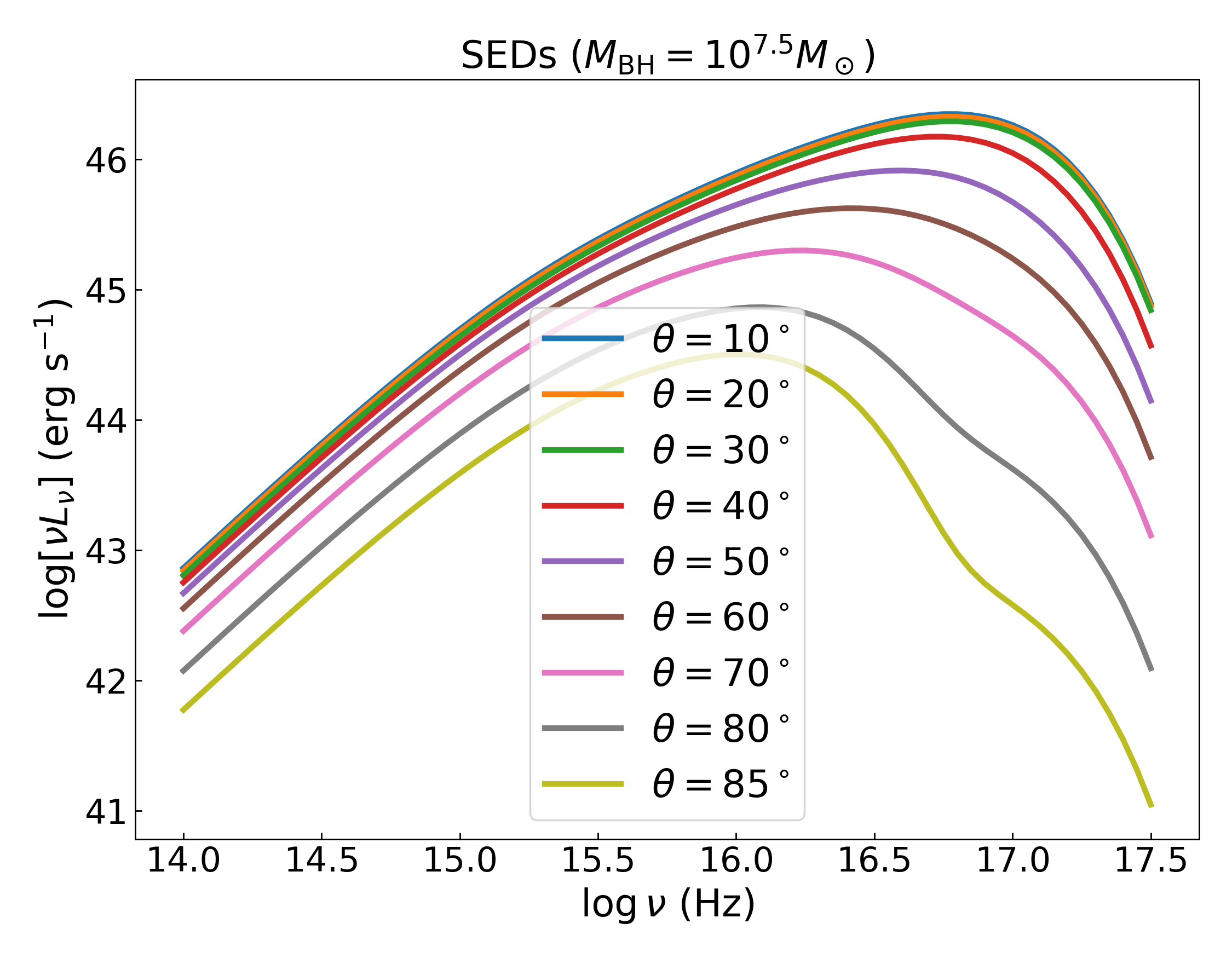}
\caption{Thick disk non-blackbody spectra for Model B. The calculations assume a black hole mass of $\Mh=10^{7.5}\,\msun$, a scattering-dominated (constant density) atmosphere with gas to total pressure ${\cal P}=10^{-4}$, and include the self-illumination effect of the funnel walls.
}
\label{fig:spectra}
\end{figure}

Figure \ref{fig:spectra} shows our theoretical SEDs as a function of the viewing angle. For this calculation we assumed a black hole mass of $\Mh=10^{7.5}\,\msun$ and a scattering-dominated (constant density) atmosphere with gas to total pressure ${\cal P}=10^{-4}$. The frequency-dependent anisotropy of the radiation field is evident; the observed soft X-ray flux ($\sim 0.4$ keV) decreases by approximately 1.0 dex as the inclination increases from $i=30^\circ$ to $i=70^\circ$. This sharp decline occurs as the innermost regions become shadowed by the thick disk geometry. Notably, even for $i>\Theta$, a fraction of this inner radiation is reprocessed and reflected by the funnel walls into the line of sight. In contrast, the observed optical flux over the same inclination range decreases by only 0.15 dex, reflecting the more isotropic distribution of the cooler, outer disk emission.

\subsection{Ultra-blue UV continuum slopes}

\begin{figure}[!htb]
\centering
\includegraphics[width=\hsize]{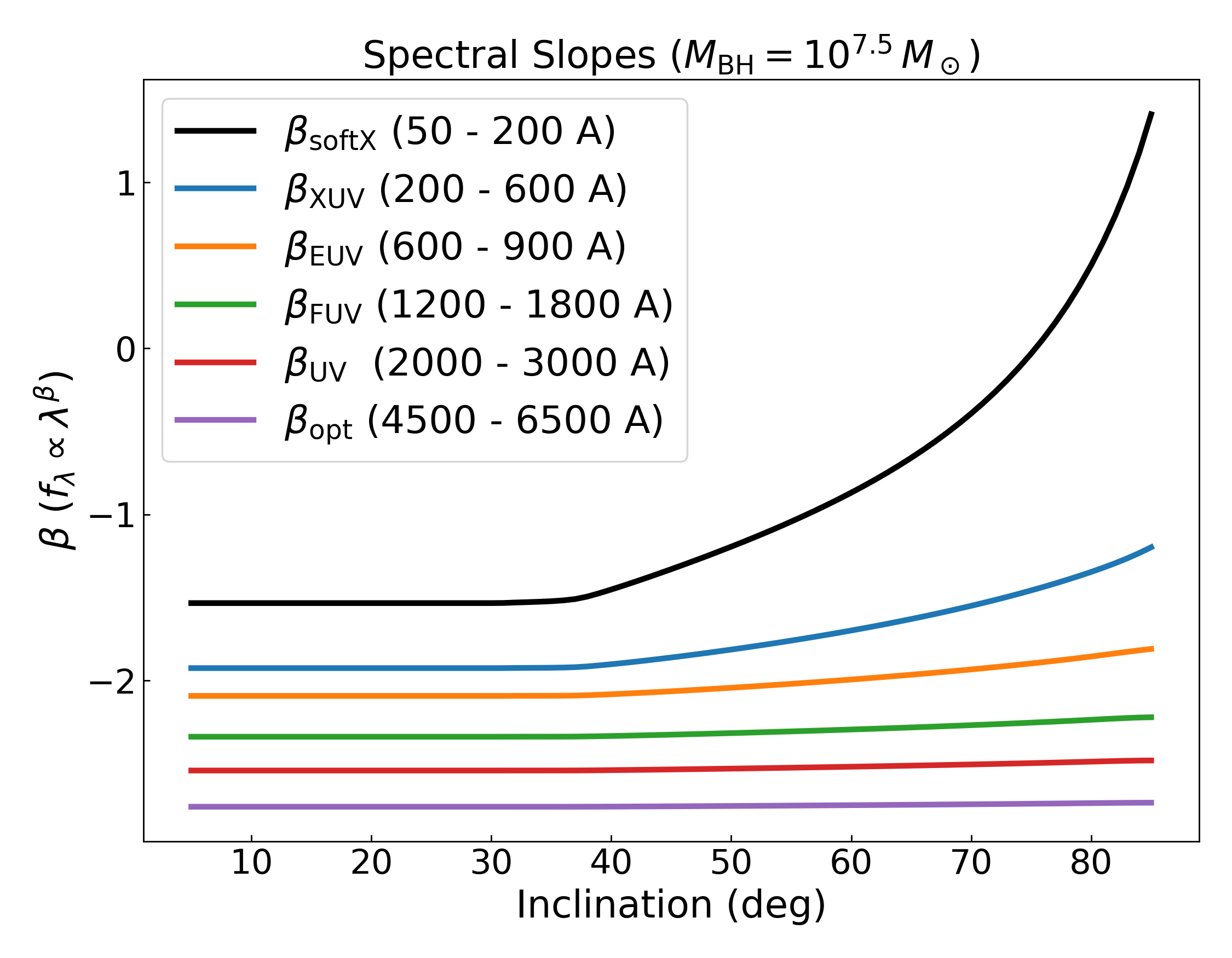}
\caption{Inclination dependence of the continuum slopes $\beta$, defined by fitting the emergent spectrum with a power law $f_\lambda(i)\propto \lambda^\beta$ over six rest-frame wavelength windows: softX (50--200\,\AA), XUV (200--600\,\AA), EUV (600--900\,\AA), FUV (1200--1800\,\AA), UV (2000--3000\,\AA), and optical (4500--6500\,\AA), as indicated in the legend. The calculations assume a fiducial super-Eddington accretor with $M_{\rm BH}=10^{7.5}\,M_\odot$ and $\dot m=12.4$ (Model B). At low-to-intermediate inclinations ($i\lesssim35^\circ$) the slopes are nearly constant, while toward edge-on views the ionizing continuum softens: $\beta_{\rm softX}$, $\beta_{\rm XUV}$, and $\beta_{\rm EUV}$ increase markedly, whereas the FUV changes only modestly and the UV and optical slopes remain nearly unchanged.
}
\label{fig:beta}
\end{figure}

Figure~\ref{fig:beta} shows the inclination dependence of the continuum slopes $\beta$, defined by fitting the angle-dependent emergent spectrum with a power law $f_\lambda(i)\propto \lambda^\beta$ over six rest-frame wavelength windows (softX, XUV, EUV, FUV, UV, and optical), as indicated in the legend. At low-to-intermediate inclinations ($i\lesssim 35^\circ$) the slopes are nearly constant, while toward edge-on views the short-wavelength continuum becomes progressively redder: $\beta_{\rm softX}$, $\beta_{\rm XUV}$, and $\beta_{\rm EUV}$ increase markedly (corresponding to a softer spectrum in $f_\nu$), whereas the FUV changes only modestly and the UV--optical slopes remain almost unchanged. This illustrates that the strongest orientation-dependent spectral shaping occurs in the ionizing bands, with comparatively weak impact on the rest-frame optical continuum.

We note that these models produce an exceptionally blue rest-UV continuum, steeper than the standard thin-disk asymptotic slope $f_\lambda\propto \lambda^{-7/3}$ ($\beta=-2.33$). Neither of these slope predictions align well with observations, where reddening by dust intrinsic to the source or host galaxy is commonly invoked to explain the overall redder slope observed in quasar spectra at low redshifts \citep{Davis2007,Capellupo2015,Baron2016}.
There is evidence that the level of dust extinction in super-Eddington AGNs may be higher than that of slow-accreting sources \citep{Castello2016}. Similarly ultra-blue UV spectra are measured in the hydrodynamical simulations of supercritical flows by \citet{Pacucci2024}. Interestingly, an extremely blue slope of $\beta=-2.94\pm 0.03$ has been measured in the quasar PSO J006+39, a candidate super-Eddington source at $z=6.62$ \citep{Tang2019}. In the $(\beta_{\rm UV},\beta_{\rm opt})$ plane, several JWST LBDs exhibit extremely blue continua, with spectral slopes approaching those expected for an unreddened accretion disk \citep[e.g.,][]{Brazzini2026}.

\subsection{Broad emission lines}

Rest-frame UV-optical emission lines are powerful diagnostics of the ionizing SED in AGNs, particularly in the far-UV, extreme-UV, and soft X-ray ranges that are inaccessible due to interstellar and intergalactic absorption \citep[e.g.,][]{Osterbrock2006}. Variations in line ratios and equivalent widths (EWs) probe the ionization parameter, metallicity, covering factor, geometry, and -- critically -- the Eddington ratio and black hole mass \citep{BorosonGreen1992,Ferland2020}. The anomalies observed in JWST BLAGNs highlight these degeneracies in line-based studies, while also suggesting that the ionizing continuum and/or the structure of the broad-line region (BLR) structure may differ substantially from those of classical AGNs.

\begin{figure*}[!htb]
\centering
\includegraphics[width=0.9\hsize]{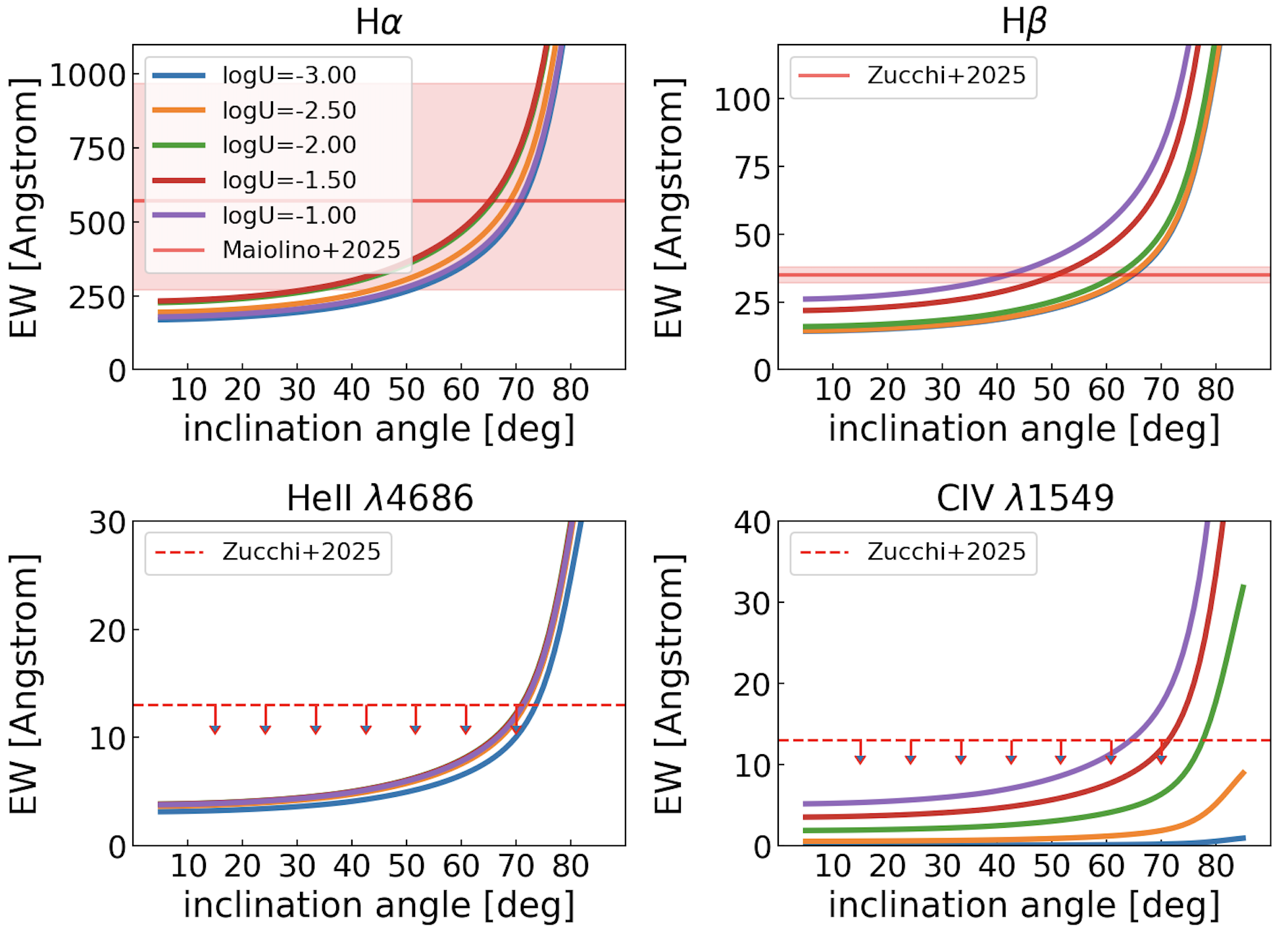}
\caption{
Predicted rest-frame EWs of broad H$\alpha$, H$\beta$, He{\sc ii} $\lambda 4686$, and C{\sc iv} $\lambda 1549$ emission lines as a function of observer inclination angle. The curves represent photoionization calculations performed with \textsc{Cloudy} using the input SED of {Model~B} (corresponding to $\dot m=12.4$) for a range of ionization parameters ($\log U = -3.0$ to $-1.0$, as labeled). The models assume a BLR gas distribution concentrated near the disk plane -- where it is shielded from the full funnel radiation field -- and a global covering factor of $C_{\rm BLR} = 15\%$. All calculations adopt a metallicity of $0.1\,Z_\odot$ and a gas density of
$\log(n_{\rm H}/{\rm cm^{-3}}) = 10$, and are stopped at a total hydrogen column of $\log(N_{\rm H}/{\rm cm^{-2}}) = 23$. The horizontal red line and shaded red region in the top left
panel indicate the median EW ($570$\,\AA) and the 95\% confidence interval for the $z>4$ BLAGN sample from \citet{Maiolino2025}. Measurements with $1\sigma$ uncertainties and $3\sigma$ upper limits for broad H$\beta$, He{\sc ii}, and C{\sc iv} are from the stacked spectra of $z>5$  JWST blue BLAGNs of \citet{Zucchi2025}. The sharp decrease in EWs at low inclination angles arises from the geometric collimation of the super-Eddington continuum: face-on observers see an isotropic-equivalent continuum that is significantly brighter than the one ionizing the BLR, resulting in the dilution of line features.
}
\label{fig:EW}
\end{figure*}

A key observational clue is that BLAGNs show abnormally strong H$\alpha$ emission: the reported median rest-frame EW of the broad H$\alpha$ line in JWST-identified $z>4$  is $\simeq 570$~\AA, roughly a factor of $\sim 3$ higher than in typical low-$z$ AGNs \citep{Maiolino2025}. Large Balmer EWs require intrinsically enhanced line emission -- for example, an EUV-bright ionizing SED and/or a large BLR covering factor. The recent stack of $z>5$ JWST-selected LBDs from the JWST Advanced Deep Extragalactic Survey (JADES) assembled by \citet{Zucchi2025} provides additional observational constraints. Although low-ionization Balmer lines remain prominent, with H$\beta$ showing an EW of $37\pm 5$ \AA\ (broad component), high-ionization emission is markedly suppressed: C{\sc iv} $\lambda 1549$ and He{\sc ii} $\lambda 4686$ are undetected 
-- lines that are nearly ubiquitous in lower-redshift Type~1 AGNs \citep[e.g.,][]{vandenberk_2001,richards_2011,Wu2022}. 

\begin{figure}[!htb]
\centering
\includegraphics[width=\hsize]{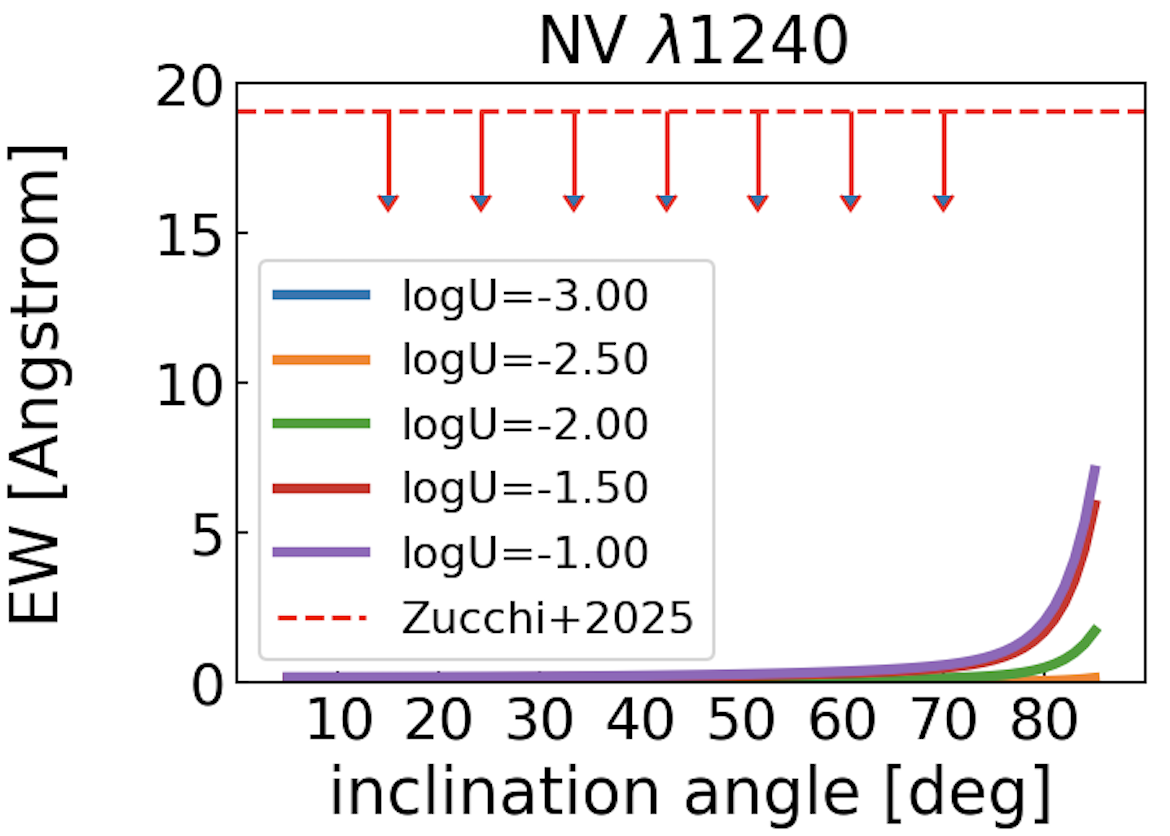}
\caption{Predicted rest-frame equivalent width of broad N{\sc v} $\lambda1240$ as a function of inclination angle, $i$, for our fiducial BLR model with sub-solar metallicity ($Z=0.1,Z_\odot$). Solid curves show predictions for different ionization parameters (legend). At $Z=0.1,Z_\odot$, broad N{\sc v} is expected to be extremely weak over most of the parameter space, with appreciable EWs arising only for large ionization parameters and near-equatorial sightlines. Therefore, any robust detection of broad N{\sc v} in such systems would likely require nitrogen enhancement and/or viewing angles close to the disk plane.
}
\label{fig:NV}
\end{figure}

Relatively weak high-ionization emission lines generally imply that the BLR is illuminated by a sufficiently soft ionizing SED. 
Although super-Eddington flows generate an intrinsically hard, UV-bright spectrum within the inner funnel, this radiation is geometrically beamed along the polar axis. Consequently, a BLR located near the equatorial plane will be illuminated by the much softer self-shielded emission of the outer flow. 
The observed EWs will be further modified by the viewing angle. While line ratios are set primarily by the SED incident on the BLR (i.e., the equatorial illumination), the EWs depend on the continuum seen by the distant observer. An observer at lower inclinations  (more face-on) sees a brighter funnel continuum, which dilutes the lines and reduces their EWs. Conversely, an observer at high inclination sees a fainter, self-shielded continuum, resulting in larger observed EWs for the same intrinsic line luminosity. To test these ideas, we constructed a grid of photoionization models tailored to conditions typical  of the BLR. All models assume a metallicity of $0.1\,Z_\odot$ (broadly consistent with estimates for BLAGNs at $z \gtrsim 5$, \citealt{maiolino2024J,trefoloni_2025}) and the ionization parameter $U$ was varied from $\log U = -3.0$ to $-1.0$. We ran separate model grids at fixed gas densities $\log(n_{\rm H}/{\rm cm^{-3}})=9$, 10, and 11.
In parallel, we compared the predicted EWs of H$\alpha$, H$\beta$, He{\sc ii} $\lambda 4686$, and C{\sc iv} $\lambda 1549$ with the observed values.

We adopt the working assumption that the BLR gas is distributed predominantly in the equatorial plane of the accretion flow. This is motivated by a range of low-redshift constraints, including dynamical reverberation-mapping inferences of an axisymmetric BLR with a finite opening angle \citep[e.g.,][]{Du2025}, as well as microlensing-induced broad-line profile distortions in lensed quasars that favor disk-like or equatorial BLR geometries over polar-wind configurations \citep[e.g.,][]{Hutsemekers2023,Savic2024}. Direct reverberation-based modeling of well-sampled light curves also favors thin disk-like BLR geometries \citep{Nunez2014}. In practice, we represent the BLR as a clumpy, equatorially concentrated distribution in which the mean number of clouds intersected by a line of sight increases rapidly toward the disk plane 
\citep{Nenkova2008}, so the probability that direct disk photons escape without absorption decreases accordingly at high inclinations. The overall normalization is set by requiring that the angle-averaged cloud-interception probability matches the global BLR covering factor.

Each cloud was truncated at $\log(N_{\rm H}/{\rm cm^{-2}})=23$, a value that  is routinely adopted in BLR photoionization calculations because it is sufficient to span the full hydrogen Strömgren layer and the extended partially ionized zone where the bulk of the UV/optical broad-line emission (e.g., the Balmer series, He{\sc ii}, and C{\sc iv}) is produced. Extending the integration to larger columns (or, equivalently, to a fixed outer radius) would primarily append a cold, predominantly neutral tail that contributes negligibly to the line equivalent widths analyzed here, while introducing additional uncertainties associated with poorly constrained physics in the far-neutral gas. We emphasize that these BLR slabs are not intended to be the dominant X-ray absorber in our model: the X-ray weakness is instead driven by the intrinsic softness and weakness of the coronal component and by geometric self-shielding in the super-Eddington thick inner flow, whose equatorial columns are expected to be Compton-thick. Any additional dusty circumnuclear material at larger radii would mainly redden and dim the observed continuum (and line fluxes) along high-inclination sightlines; it would not change the intrinsic BLR line production, and it would leave the observed EWs unchanged if it attenuates the line and adjacent continuum similarly.

For a fiducial $M_{\rm BH}=10^{7.5}\,M_\odot$ accretor with $\dot m=12.4$, adopting the equatorial ($i=85^\circ$) incident SED appropriate for BLR illumination
($Q_{\rm H}=10^{55}\,{\rm s^{-1}}$), the range $\log U\in[-3.0,-1.0]$ corresponds to BLR distances $\simeq 2\times10^{17}$ to $2\times10^{18}\,{\rm cm}$.

\begin{figure}[!htb]
\centering
\includegraphics[width=\hsize]
{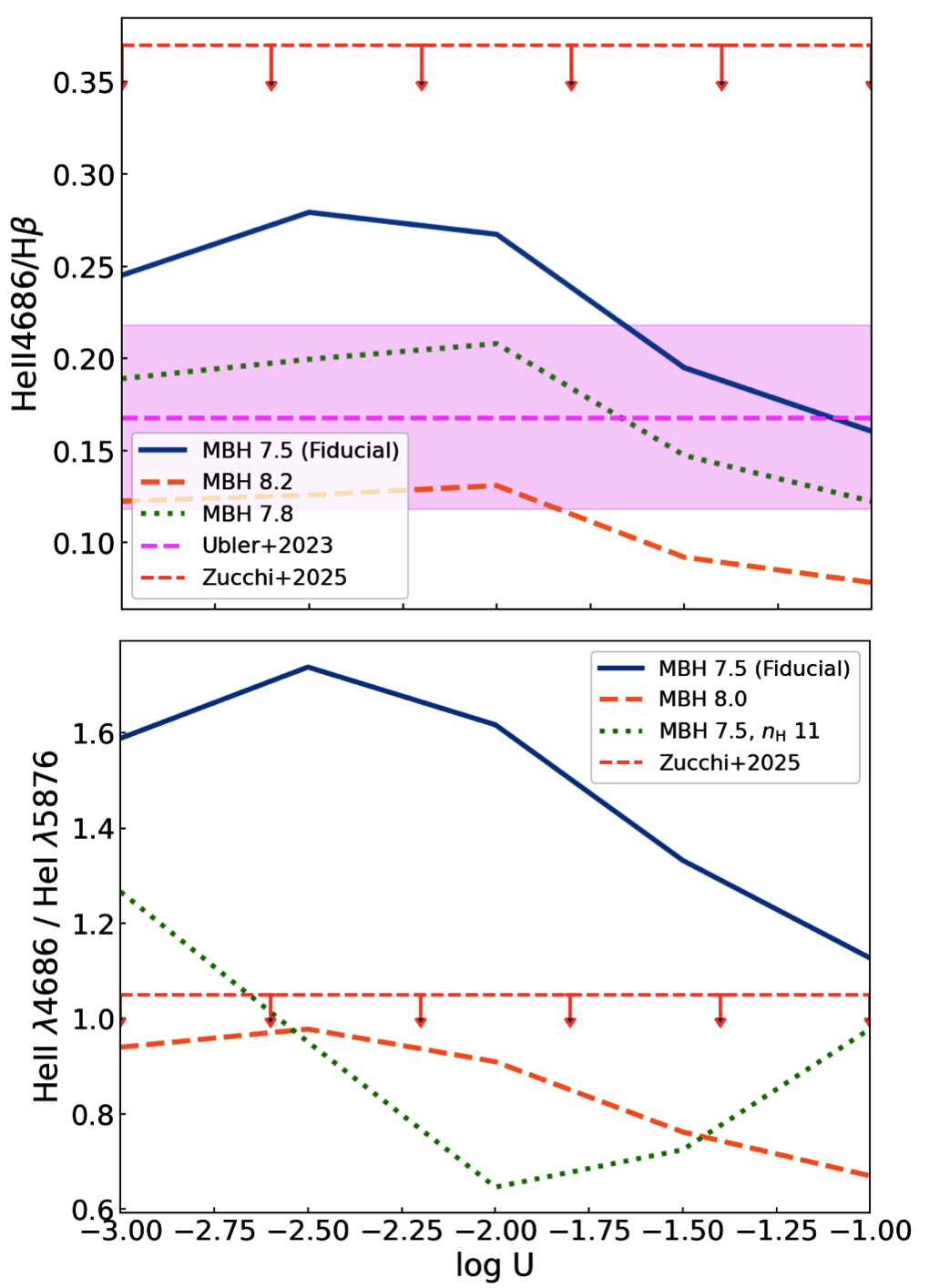}
\caption{Predicted broad-line ratios 
 as a function of ionization parameter, $\log U$, from our {\sc Cloudy} grid. The curves illustrate the sensitivity of these diagnostics to the hardness of the incident continuum, which becomes softer for larger black-hole masses, and to the gas density. All calculations adopt the same BLR gas parameters as in Figure~\ref{fig:EW} (metallicity $0.1\,Z_\odot$, density $\log(n_{\rm H}/{\rm cm^{-3}})=10$, and total hydrogen column $\log(N_{\rm H}/{\rm cm^{-2}})=23$). Top panel: He{\sc ii} $\lambda4686$/H$\beta$. The solid, dotted, and dashed curves correspond to models with $\log(M_{\rm BH}/M_\odot)=7.5$ (fiducial), 7.8, and 8.2, respectively. For comparison, the dashed magenta line and shaded magenta band show the value reported by \citet{Ubler2023} for the $z=5.55$ AGN \texttt{GS\_3073} (which has an estimated black-hole mass of $\log(M_{\rm BH}/M_\odot)=8.2\pm0.4$), while the dashed red line indicates the upper limit from the stack of LBDs \citep{Zucchi2025} (downward arrows). Bottom panel: He{\sc ii} $\lambda4686$/He{\sc i} $\lambda5876$. The solid, dotted, and dashed curves correspond to models with 
$(n_{\rm H},M_{\rm BH})=(10^{10}\,{\rm cm}^{-3},10^{7.5}\,M_\odot), (10^{10}\,{\rm cm}^{-3},10^8\,M_\odot),$
and $(10^{11}\,{\rm cm}^{-3},10^{7.5}\,M_\odot)$, 
respectively.
}
\label{fig:heii_hbeta_comparison}
\end{figure}

\subsubsection{Strong Balmer lines at low covering factors}

Figure~\ref{fig:EW} illustrates how the dependence on inclination of predicted equivalent widths provides a natural  explanation for the JADES line measurements. Although the model curves fall within the broad 95\% distribution of the H$\alpha$ observations across a wide range of viewing angles, simultaneously reproducing the median H$\alpha$ value and the stacked measurements for H$\beta$, He{\sc ii}, and C{\sc iv} requires the typical LBD to be viewed at intermediate inclinations,
$50^\circ$--$70^\circ$, a regime in which the continuum  is partially de-boosted relative to face-on orientations. In the same inclination  range, the predicted EWs of He{\sc ii}   and C{\sc iv} remain below the observed upper limits of $\lta 13$--14\,\AA, while rapidly exceeding 
these limits at higher inclinations. Thus, a single, moderate inclination of  $\sim 60^\circ$ simultaneously accounts for the large Balmer EWs and the marked  weakness of the high-ionization lines. This inclination range is also  consistent with expectations for a random Type~1 AGN viewed within the dusty torus  opening angle in unified models and matches the broad inclination distributions  inferred from BLR dynamical modeling and orientation indicators. The model therefore captures key spectral properties of the JADES $z>5$ LBDs  without fine tuning. Notably, because inherently blue disk spectra provide a much higher ratio of ionizing to optical photons than standard quasar composites, the observed large Balmer EWs are matched with modest covering factors, $C_{\rm BLR} \sim 10$--$20$\% (we adopt 15\% in Fig. \ref{fig:EW} for illustration), without invoking enshrouded geometries \citep[][]{Inayoshi2025}. In our scenario, the most extreme H$\alpha$ emitters are not necessarily those with the largest covering factors, but rather those viewed at high inclination. 

\subsubsection{N{\sc v} $\lambda$1240}

Producing N{\sc v} requires photons above 77.5 (N{\sc iv}$\rightarrow$N{\sc v}), while photons above 97.9 eV (N{\sc v}$\rightarrow$N{\sc vi}) efficiently deplete it. In our sub-solar abundance BLR grids, broad N{\sc v} $\lambda1240$ is generically weak. Figure \ref{fig:NV} shows that the predicted rest-frame EW is $\lta$ a few angstroms over most inclinations, rising steeply only for near-equatorial sightlines at high ionization parameter. The upturn at large $i$ primarily reflects the strong inclination-dependent suppression of the observed UV continuum, which increases the EW even when the intrinsic line emissivity changes more modestly.
Consequently, a clear detection of this line in LBDs would not be expected at sub-solar abundances under the baseline assumption of scaled-solar abundance ratios. 
If broad N{\sc v} is reported at high significance, it likely indicates enhanced N/H (and/or N/O) relative to scaled-solar values -- consistent with nitrogen behaving as a secondary element at high metallicity -- rather than being a generic outcome of photoionization at low $Z$.

\subsubsection{He{\sc ii} $\lambda4686$/H$\beta$}

Line ratios provide a more direct probe of the detailed shape of the ionizing continuum and of the physical conditions in the emitting plasma. Figure~\ref{fig:heii_hbeta_comparison} shows the {\sc Cloudy}-predicted He{\sc ii} $\lambda4686$/H$\beta$ ratio as a function of ionization parameter over the range explored in our grid, for fixed BLR gas parameters (metallicity, density, and stopping column) adopted elsewhere in this work. The predicted ratio depends on $\log U$, but is also sensitive to the hardness of the extreme-UV continuum that controls the supply of photons above the He$^{+}$ edge. In our super-Eddington model family, increasing $M_{\rm BH}$ softens the incident SED and lowers He{\sc ii} $\lambda4686$/H$\beta$, bringing the predictions into close agreement with the observational constraints for the $z=5.55$ BLAGN \texttt{GS\_3073}, which provides one of the clearest detections of broad He{\sc ii} $\lambda4686$ at high redshift \citep{Ubler2023}. For this source, \citet{Ubler2023} estimate $\log(M_{\rm BH}/M_\odot)=8.2\pm0.4$ from single-epoch virial scaling relations using the luminosity and FWHM of the broad H$\alpha$ line. For reference, \citet{Wu2022} measure a broad He{\sc ii} $\lambda4686$/H$\beta$ ratio of $\simeq 0.2$ in a large SDSS quasar sample dominated by luminous objects, with a black-hole mass distribution peaking at $\log(M_{\rm BH}/M_\odot)\simeq 8.7$--$8.9$. 

For context, \citet{Ferland2020} predict that He{\sc ii}~$\lambda4686$/H$\beta$ increases with Eddington ratio, from $\simeq 0.17$ to $\simeq 0.49$ in their \textsc{Cloudy} BLR calculations (Table~1); the corresponding Case-B photon-counting estimates span $\simeq 0.29$ to $\simeq 0.77$. By shielding the equatorial BLR from the funnel, our models predict substantially reduced He{\sc ii} $\lambda4686$/H$\beta$ ($\simeq 0.10$--$0.25$), well below the hard super--critical case.

\subsubsection{He{\sc ii} $\lambda4686$/He{\sc i} $\lambda5876$}

The lower panel of Figure~\ref{fig:heii_hbeta_comparison} shows the predicted He{\sc ii} $\lambda4686$/He{\sc i} $\lambda5876$ ratio. Because both lines are helium recombination features, this ratio traces the He$^{++}$/He$^{+}$ ionization balance and is sensitive to the relative supply of photons above 54.4~eV versus 24.6~eV \citep[e.g.,][]{Richardson2014}. However, He{\sc i} $\lambda5876$ is a triplet line whose emissivity depends on density and optical-depth effects, making the ratio particularly sensitive to $n_{\rm H}$ within our BLR framework. The upper limit inferred from the LBD stack of \citet{Zucchi2025} lies below the predictions of our fiducial $n_{\rm H}=10^{10}\,{\rm cm^{-3}}$ and 
$\log(M_{\rm BH}/M_\odot)=7.5$, 
model over the full $\log U$ range. This tension can be alleviated by higher densities ($n_{\rm H}\sim10^{11}\,{\rm cm^{-3}}$) and/or larger black hole masses, where the enhanced He{\sc i} emission drives the ratio down into agreement with the stacked constraint.

\begin{figure}[!htb]
\centering
\includegraphics[width=\hsize, trim=0 1.mm 0 1.mm,clip]{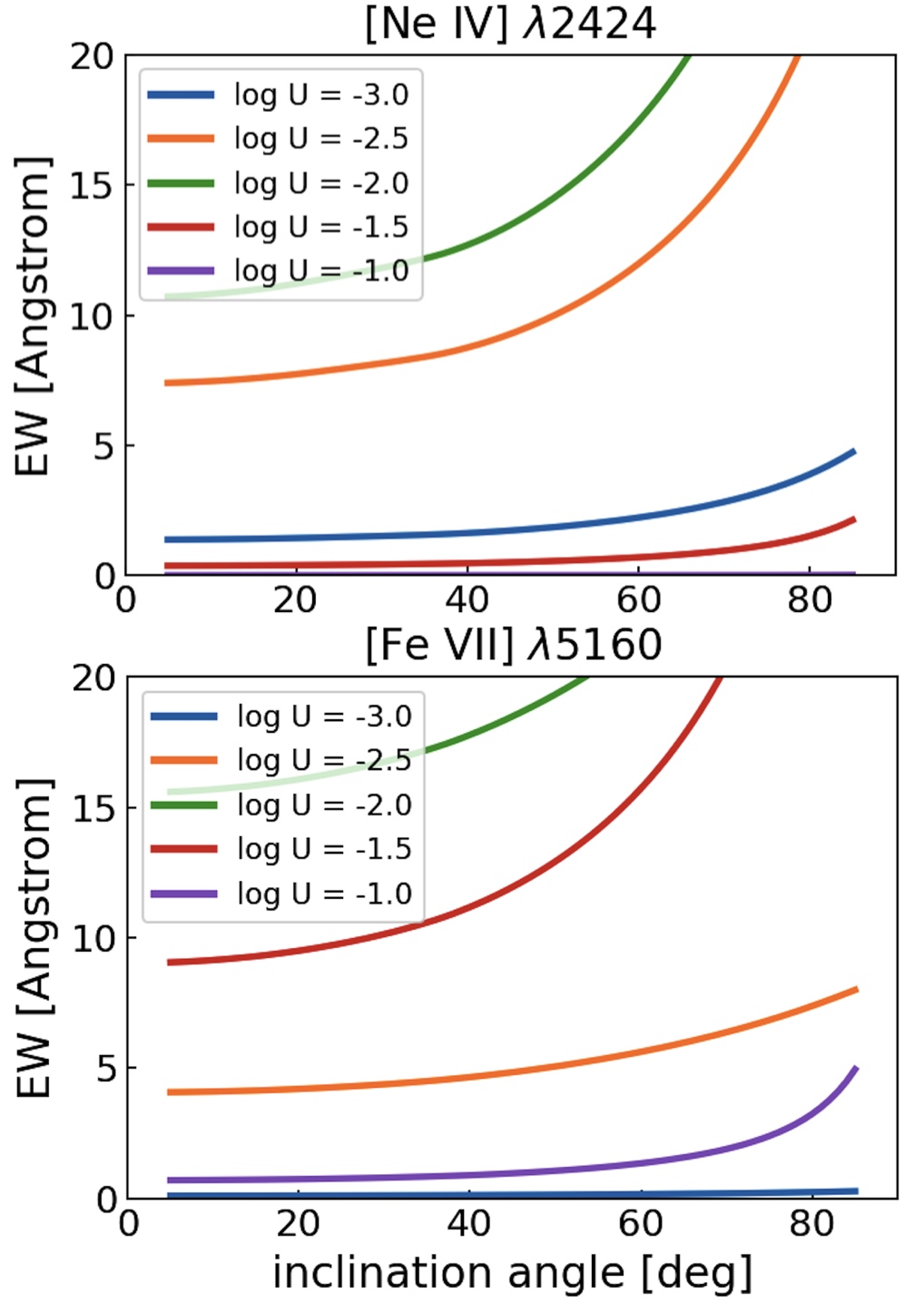}
\caption{Predicted rest-frame EWs of the high-ionization lines [Ne{\sc iv}] $\lambda$2424 (top) and Fe{\sc vii} $\lambda$5160 (bottom) as a function of observer inclination angle. The curves represent {\sc Cloudy} photoionization models for an NLR or CLR with gas density $n_{\rm H}=10^5$ cm$^{-3}$, column density $N_{\rm H}=10^{22}$ cm$^{-2}$, metallicity $0.1\,Z_\odot$, and a global covering factor of $C_{\rm NLR}=10$\%. The clouds are assumed to be located along the polar axis and illuminated by the hard, unattenuated funnel continuum (modeled using the Model B SED at $i=30^\circ$). Different colors correspond to ionization parameters ranging from $\log U = -1.0$ to $-3.0$. The rise in EWs at high inclinations ($i \gtrsim 50^\circ$) results from the geometric suppression of the observed continuum flux due to self-shadowing, while the intrinsic line luminosity remains high, driven by the intense polar searchlight.}
\label{fig:NeIV}
\end{figure}

\subsection{The polar searchlight: high-excitation lines}
\label{sec:polar_searchlight}

A consequence of the thick-disk funnel geometry is the extreme anisotropy of the ionizing radiation field in the soft X-ray/EUV regime. As illustrated by the black curve in Fig.~\ref{fig:beta}, the spectral slope $\beta_{\rm softX}$ (covering the 50--200\,\AA\ range) exhibits a steep dependence on inclination. Observers viewing the system near the symmetry axis ($i \lesssim 30^{\circ}$) see a hard, rising continuum with $\beta_{\rm softX} \approx -1.5$. However, this slope transitions to $\beta_{\rm softX} > 0$ at inclinations $i \gtrsim 70^{\circ}$, since the innermost, hottest regions of the funnel are geometrically self-shielded by the torus walls. This geometrical beaming is even more pronounced in flows accreting at higher rates (e.g., Model A), where the funnel is narrower and the radiation field is more strongly collimated.
This spectral ``switch'' may have  profound implications for the spatial distribution of high-ionization species. The 50--200\,\AA\ window ($60$--$250$\,eV) is physically critical as it spans the ionization potentials that set high-ionization narrow UV lines and coronal lines, including the UV line [Ne{\sc iv}] $\lambda$2424 ($63$ eV), the permitted narrow UV line N{\sc v} $\lambda$1240 (when present; $77$ eV), and optical coronal iron lines such as [Fe{\sc vii}] $\lambda$5160 ($100$ eV).

In our super-Eddington framework, gas located in the equatorial plane -- or at high inclinations generally -- sees a ``dark'' sky in this energy band, effectively suppressing the formation of these species. In contrast, gas clouds located along the symmetry axis are bathed in an intense searchlight of hard ionizing photons. Consequently, we predict that super-Eddington BLAGNs should exhibit a biconical ionization structure: while the equatorial BLR may be deficient in high-ionization emission due to shielding, the polar narrow-line region (NLR) and coronal line region (CLR) should be highly excited.
This geometric stratification may help interpret the minority of high-$z$ sources showing detectable high-ionization UV lines.  \citet{Tang2025} note that only four LRDs currently have grating spectra deep enough to detect narrow N{\sc v} at low EW, and they detect narrow N{\sc v} in CEERS-7902. Similarly, the detection of strong high-ionization lines such as [Ne{\sc iv}] $\lambda$2424 in GN-z11 ($z=10.6$, \citealt{Maiolino2024Nat}) and the tentative detection of [Fe{\sc vii}] $\lambda$5160 in THRILS 46403 ($z=6.58$, \citealt{Lambrides2025}) may support the presence of a hard ionizing source capable of stripping inner-shell electrons, consistent with a face-on view of a super-Eddington funnel. 

We have explored the observability of specific high-ionization transitions by modeling a polar NLR or CLR illuminated by the hard funnel continuum. Figure~\ref{fig:NeIV} presents the predicted equivalent widths for the UV [Ne{\sc iv}] $\lambda$2424 doublet and Fe{\sc vii} $\lambda$5160 as a function of inclination. The detectability of these lines depends strongly on the ionization parameter, though the response is not strictly monotonic.
For [Ne{\sc iv}] $\lambda$2424, the emission peaks at intermediate ionization, with the largest equivalent widths obtained for $\log U \simeq -2.5$ to $-2.0$. In this regime the line becomes prominent, reaching EWs of $10$--$20\,$\AA\ at high inclinations where the primary continuum is suppressed by shadowing. At lower ionization parameters ($\log U \lesssim -3.0$) the line rapidly fades to EW $\lesssim 2\,$\AA, while at the highest ionization shown ($\log U=-1.0$) the predicted EW is again extremely small, consistent with over-ionization of Ne$^{3+}$.

Although a marginal emission feature is observed near rest-frame 5160\,\AA\ in THRILS~46403 \citep{Lambrides2025}, our photoionization modeling disfavors its identification as [Fe{\sc vii}] $\lambda5160$.  In particular, as already emphasized by Lambrides et al., a detection of $\lambda5160$ without a corresponding detection of the stronger [Fe~{\sc vii}] $\lambda6087$ line is difficult to reconcile with standard coronal-line physics. As demonstrated in Figure~\ref{fig:NeIV}, the predicted EW of [Fe{\sc vii}] is strongly sensitive to the ionization parameter. At the extremes of our grid ($\log U=-1.0$ and $-3.0$), [Fe~{\sc vii}] $\lambda5160$ remains weak (EW $\lesssim$ a few \AA\ even at high inclinations). For intermediate ionization parameters ($\log U\simeq -2.5$ to $-1.5$),
[Fe{\sc vii}] $\lambda5160$ is predicted to be readily detectable: in this regime, however, the accompanying [Fe~{\sc vii}] $\lambda6087$ transition should be substantially stronger.  Given the non-detection of [Fe{\sc vii}] $\lambda6087$, we conclude that the physical conditions required for genuine [Fe{\sc vii}] emission are unlikely to be present, and the 5160\,\AA\ feature is more plausibly a statistical fluctuation or an unidentified blend.

The predicted EWs scale approximately linearly with the NLR or CLR covering factor; thus, for $C_{\rm NLR} \lesssim 5\%$ the [Ne{\sc iv}] and [Fe{\sc vii}] features would be weakened by a factor $\gtrsim 2$ relative to Figure~\ref{fig:NeIV} and may fall below current detectability in typical JWST spectra. Thus, the detection of these species serves as a dual diagnostic: it selects sources viewed at sufficiently high inclinations to suppress the continuum (enhancing line contrast) and requires highly ionized gas with non-negligible covering factor, photoionized by the polar searchlight.

\section{Conclusions}

We have examined super-Eddington accretion flows as an explanation for several puzzling properties of the high-$z$ LBD population uncovered by JWST. Using geometrically thick, non-advective disk solutions with self-illumination and inclination-dependent radiative transfer, we find:

\begin{itemize}

\item Super-Eddington accretion generically produces a geometrically thick torus ($h/r>1$) with a narrow polar funnel. The emergent radiation field is strongly anisotropic: the isotropic-equivalent luminosity is largest along the polar axis, while self-shadowing suppresses the flux seen at high inclinations.

\medskip
\item The torus walls emit a hot, modified-blackbody continuum that yields very blue rest optical-UV slopes, with $\beta \simeq -2.6$.

\medskip
\item The same anisotropy helps reconcile strong Balmer emission with weak high-ionization broad lines. A BLR concentrated near the disk plane is illuminated predominantly by the equatorial, spectrally filtered continuum and is effectively shielded from the hardest funnel radiation, reducing He {\sc ii} and C {\sc iv} production. Large Balmer EWs can be produced with modest covering factors because the BLR-illuminating continuum has a higher ionizing-to-optical photon budget than standard quasar composites, while the observer’s optical/UV continuum is inclination-dependent (boosted for face-on views and diluted at higher inclinations), without invoking an enshrouded geometry. 

\medskip 
\item The precedent point concerns BLR production rather than line-of-sight visibility: narrow high-ionization lines, such as [Ne{\sc iv}] $\lambda$2424, can still arise in more extended gas that intercepts some fraction of the escaping polar radiation.

\medskip
\item Finally, the same funnel geometry offers a natural route to the observed X-ray weakness of many JWST BLAGNs. In a reflection-dominated, funnel-like configuration the coronal non-thermal hard X-ray component is intrinsically weak, while the thermal soft X-ray emission is efficiently blocked along equatorial and intermediate sightlines by the torus and becomes visible only for more poleward observers.

\end{itemize}

Taken together, these results  may support the idea that a significant fraction of JWST-selected LBDs is powered by rapidly growing black holes accreting at supercritical rates, with orientation-dependent continua and line diagnostics set by funnel collimation and equatorial shielding. In future work we will assess whether the LRD-LBD phenomenology is driven primarily by inclination within a common engine, as in our anisotropic supercritical models, or instead reflects an evolutionary sequence in which the obscuring column declines with time and progressively unveils a bluer intrinsic SED, as proposed in the direct-collapse scenario of \citet{Pacucci2026}.

\begin{acknowledgements}

Support for this work was provided by NASA through grant TCAN 80NSSC21K027 and  by grant NSF PHY-2309135 to the Kavli Institute for Theoretical Physics (KITP). We acknowledge many useful discussions on this project with R. Maiolino and thank G. Zucchi for providing line-intensity measurements and for guidance with {\sc Cloudy} modeling.

\end{acknowledgements}

\bibliographystyle{aa}
\bibliography{paper}

\begin{appendix}
\section{Shadowing}

We revisit the shadowing formalism of \citet[]{Madau1988}, extended to account for the curvilinear limb of the funnel outer rim that becomes relevant at high inclinations. The observed flux at distance $D$ is
\begin{equation}
F = \frac{1}{D^{2}}\int_{\Sigma_{\rm vis}} I(\hat n)\,(\hat n\cdot \hat N)\, d\Sigma,
\label{eq:Fobs}
\end{equation}
where $\hat N$ is the outward unit normal, $\hat n$ is the line-of-sight unit vector (from the surface element toward the observer), and $\Sigma_{\rm vis}$ is the portion of the emitting surface not obscured by the disk itself. 
We adopt a Cartesian frame $(x,y,z)$ centered on the black hole, with the line of sight in the $xz$-plane at inclination $i$ to the $z$ axis,
\begin{equation}
\hat n=(\sin i,\,0,\,\cos i).
\label{eq:hatn}
\end{equation}
A point on the surface $z=h(r)$ is $P=(r\cos\varphi,\,r\sin\varphi,\,h)$, and the corresponding unit normal (choosing the outward, $z$-increasing branch) is
\begin{equation}
\hat N=\frac{\left(-\cos\varphi\,h'(r),\,-\sin\varphi\,h'(r),\,1\right)}
{\sqrt{1+h'(r)^{2}}}.
\label{eq:hatN}
\end{equation}

\medskip
\noindent{\it Local visibility.} A surface element can contribute to the observed flux only if it faces the observer,
\begin{equation}
\hat n\cdot \hat N>0 \quad \Longleftrightarrow \quad \cos i-h'(r)\cos\varphi\,\sin i>0 .
\label{eq:local_vis}
\end{equation}

\medskip
\noindent{\it Global shadowing.} Even if Eq.~(\ref{eq:local_vis}) holds, the line of sight may intersect another portion of the funnel surface. To formalize this, we distinguish between the inner funnel ($r_{\rm in} < r < r_{\rm max}$, where $h'>0$) and the outer hump ($r > r_{\rm max}$, where $h'<0$). For the geometries considered here, global shadowing is relevant only for points on the far-side inner funnel. Introduce sky-plane coordinates $(X,Y,Z)$ obtained by a rotation about the $y$ axis,
\begin{equation}
X=\sin i\,x+\cos i\,z,\quad Y=y,\quad Z=-\cos i\,x+\sin i\,z ,
\end{equation}
with $X$ increasing toward the observer. For a given $Y$-slice of the projected image, define the local horizon by the tangent point $P_0(Y)$ on the surface such that
\begin{equation}
r_0\sin\varphi_0 = Y,
\quad
\hat n\cdot \hat N\big|_{(r_0,\varphi_0)} = 0
\ \ \Longleftrightarrow\ \ 
h'(r_0)\cos\varphi_0=\cot i .
\label{eq:tangent_conditions}
\end{equation}
Let $Z_0(Y)$ and $X_0(Y)$ be the projected height and line-of-sight coordinate of this tangent point (often located near $r_{\rm max}$). The shadowing condition applies to the far side of the inner funnel ($X<X_0$); points with $X>X_0$ lie in the foreground and are not occulted by the surface at the same $Y$. Thus, a point $P=(r,\varphi)$ is globally visible if it lies on the outer hump, or if it lies on the inner funnel and satisfies
\begin{equation}
h(r)-h(r_0) > \left[r\cos\varphi-r_0\cos\varphi_0\right]\cot i,
\label{eq:global_shadow}
\end{equation}
where the right-hand side is the height of the shadow ray cast from $P_0$ toward the observer evaluated at the $x$-coordinate of $P$, and $(r_0,\varphi_0)$ is determined by Eq.~(\ref{eq:tangent_conditions}) at the same impact parameter $Y=r\sin\varphi$. We note that the full curvilinear horizon defined here differs from a planar approximation (constant $r_0$) primarily in the outer wings of the projected image (large $|Y|$). In these regions, however, $\hat n\cdot\hat N$ is typically small because the surface becomes nearly tangent to the line of sight. Consequently, the deviations introduced by the planar approximation are negligible, affecting the integrated fluxes by only a few percent for the parameters considered here.

\section{Self-illumination effect}

Photons emitted within the funnel can scatter multiple times before escaping or being captured. Each surface element therefore receives nonlocal radiation from other visible elements, and this self-illumination must be included in the normal force balance \citep{Sikora1981,Madau1988}. We assume an empty funnel in which incident radiation is scattered isotropically (no absorption and no change in the local thermal state). For an isotropic phase function, the scattered intensity associated with an incident normal flux is \(F_{\rm in}^N/\pi\). The normal radiative equilibrium condition is
\begin{equation}
F_{\rm rad}(r) = F_{\rm rad}^{\rm local}(r) + F_{\rm in}^N(r),
\label{eq:Fradin}
\end{equation}
with the locally generated (net) flux
\begin{equation}
F_{\rm rad}^{\rm local}(r) = \frac{c}{\kappa_{\rm es}}\, g_{\rm eff}(r),
\label{eq:Fradlocal}
\end{equation}
where \(F_{\rm rad}^{\rm local}\) is the net outward flux required to balance the effective gravity normal to the surface. In the perfect-reflection limit, incident inward radiation is immediately re-emitted without thermalization, so the local temperature is set by \(F_{\rm rad}^{\rm local}\) alone, while the {total} outward flux \(F_{\rm rad}\) can exceed the local Eddington value due to nonlocal back-illumination. The normal component of the incoming scattered flux is
\begin{equation}
F_{\rm in}^N(r) = \frac{1}{\pi} \int F_{\rm rad}(r')\,
\frac{\hat{N}(r)\cdot \vec{L}}{|\vec{L}|}\, d\Omega,
\label{eq:Fin}
\end{equation}
where \(\vec{L}\) is the displacement from \(r'\) to \(r\), \(\hat{N}(r)\) is the local surface normal, and the integral extends over the solid angle subtended by the {visible} portion of the funnel wall (including shadowing). We discretize the funnel surface into concentric rings and solve the linear system
\begin{equation}
F_{{\rm rad}, i} = F_{{\rm rad}, i}^{\rm local} + \frac{1}{\pi} \sum_j B_{ij}\, F_{{\rm rad}, j},
\label{eq:Fi}
\end{equation}
with
\begin{equation}
B_{ij} = \int_{\varphi}
\frac{|(\hat{N}_i \cdot \vec{L})(\hat{N}_j \cdot \vec{L})|}{|\vec{L}|^4}\,
d\Sigma_j(\varphi),
\end{equation}
and
\begin{equation}
d\Sigma_j = r_j \left[1 + \left( \frac{dh}{dr} \right)^2_j \right]^{1/2} \Delta r\, d\varphi.
\end{equation}
The azimuthal integral is restricted to visible surface elements using the shadowing condition (Eq.~A7). 
Although the local emergent flux $F_{\rm rad,i}$ can exceed the local Eddington value because of nonlocal back-illumination, the solution of Equation~(\ref{eq:Fi}) does not create additional luminosity: it conserves the surface-integrated radiative power and only redistributes the locally generated emission over the solid angle through scattering, so the global radiative efficiency set by $L_{\rm gen}=L_{\rm rad}$ is unchanged.

We use 1000 rings and solve Eq.~\eqref{eq:Fi} via LU decomposition. The assumption that the incident field does not thermalize is appropriate when \(\kappa_{\rm es} \gg \kappa_{\rm abs}\). Once \(F_{\rm rad}(r)\) is obtained, we compute the frequency-dependent emergent intensity as
\begin{equation}
I_\nu(r) = I_\nu^{\rm local}\!\left[T_s(r)\right] + \frac{F_{\nu,\rm in}^N(r)}{\pi},
\end{equation}
where \(I_\nu^{\rm local}\) is set by the locally generated flux (and thus \(T_s\)), while \(F_{\nu,\rm in}^N\) is evaluated with the same geometry and shadowing as Eq.~\eqref{eq:Fin}, but at each \(\nu\). For each frequency, this again yields a linear system for \(I_\nu(r)\). The observed SED at a given inclination is obtained by integrating \(I_\nu(r)\) over the visible funnel surface. 
This treatment enforces force balance only in the normal direction; tangential radiation forces can accelerate surface layers \citep{Sikora_W1981,Narayan1983}. For the mildly super-Eddington regime considered here ($\dot m \sim 5$--$30$), simulations indicate that funnel regions are optically thin and the kinetic power of outflows is a small fraction of the radiative output \citep{Jiang2014,Jiang2019}. We also adopt a pseudo-Newtonian potential and neglect GR photon geodesics and capture; these effects are expected to matter most in the innermost funnel regions and are deferred to future work.

\section{Modified blackbody spectrum}

The locally produced SED of the funnel walls is approximated by the modified blackbody appropriate for isotropic scattering in a constant-density, semi-infinite atmosphere \citep{Rybicki1986,Czerny1987}:
\begin{equation}
I_\nu^{\rm local}=
\frac{2\,B_\nu(T_s)}
     {1+\sqrt{1+\kappa_{\rm es}/\kappa_{\nu,{\rm abs}}}},
\label{eq:IMBB}
\end{equation}
where \(B_\nu\) is the Planck function, \(T_s(r)\) is the surface temperature, and
\(\kappa_{\nu,{\rm abs}}\) is the (free--free + bound--free) absorption opacity. In our empty-funnel limit, incident radiation is reflected without thermalization, so \(T_s\) is set by the locally generated net flux:
\(\,cg_{\rm eff}/\kappa_{\rm es}=\pi\!\int I_\nu^{\rm local}(T_s,\rho_s)\,d\nu\). Following \citet{Hall2018}, we approximate
\begin{equation}
\kappa_{\nu,{\rm abs}}=\kappa_*\,\rho_s\,T_s^{-7/2}\,x^{-3}(1-e^{-x}),
\end{equation}
with \(x\equiv h\nu/kT_s\) and
\begin{equation}
\kappa_*=
(3.68\times10^{22}\,{\rm cm^5\,g^{-2}\,K^{7/2}})
\,[X+Y+1180Z f(T_s)](1+X),
\end{equation}
where \(X,Y,Z\) are the H, He, and metal mass fractions and \(f(T_s)\) is the neutral-metal fraction \citep{Meier2012}. We adopt \(X=0.75\), \(Y=0.248\), \(Z=0.002\), and \(f(T_s)=0.5\), giving \(\kappa_*=1.4\times10^{23}\,{\rm cm^5\,g^{-2}\,K^{7/2}}\); the electron-scattering opacity is \(\kappa_{\rm es}=0.2(1+X)\,{\rm cm^2\,g^{-1}}\). Assuming a constant ratio of gas to total pressure \({\cal P}\) in the surface layer,
\begin{equation}
\rho_s=\left(\frac{a_B\mu m_{\rm H}}{3k}\right)
\left(\frac{{\cal P}}{1-{\cal P}}\right)T_s^3,
\end{equation}
with \(\mu=0.59\) for fully ionized, low-metallicity gas. For the parameters of interest, \(\kappa_{\nu,{\rm abs}}\ll\kappa_{\rm es}\) over the UV--optical band relevant here; absorption becomes important only at \(x\lesssim0.01\) (far-IR/radio). In this limit,
\begin{equation}
I_\nu={\cal I_\nu}\,T_s^{11/4}\left(\frac{{\cal P}}{1-{\cal P}}\right)^{1/2}
\frac{x^{3/2}}{e^x(1-e^{-x})^{1/2}},
\end{equation}
where
\({\cal I_\nu}\equiv 4(c^2h^2)^{-1}\sqrt{(k^5\kappa_*/3\kappa_{\rm es})a_B\mu m_{\rm H}}
=7.2\times10^{-16}\)
erg s\(^{-1}\) cm\(^{-2}\) Hz\(^{-1}\) sr\(^{-1}\) K\(^{-11/4}\). The corresponding bolometric modified-blackbody flux is
\begin{equation}
\pi I={\cal F}\,T_s^{15/4}\left(\frac{{\cal P}}{1-{\cal P}}\right)^{1/2},
\label{eq:mbb}
\end{equation}
with \({\cal F}\equiv 1.51\,\pi{\cal I_\nu}k/h
=7.1\times10^{-5}\,{\rm erg\,s^{-1}\,cm^{-2}}\,{\rm K^{-15/4}}\). Equating Eq.~\eqref{eq:mbb} with Eq.~(\ref{eq:Frad}) yields \(T_s\gg T_{\rm eff}\) in the inner regions, extending the spectrum to higher frequencies than a sum of blackbodies. The dependence on \({\cal P}\) is weak, \(T_s\propto{\cal P}^{-2/15}\).

\end{appendix}

\label{lastpage}

\end{document}